\PassOptionsToPackage{switch}{lineno}
\documentclass{aa}
\usepackage{csvsimple-l3}
\usepackage{hyperref}
\usepackage{booktabs}
\usepackage{lscape}      
\usepackage{graphicx}
\usepackage{txfonts}
\usepackage{etoolbox}
\usepackage{xstring}  

\newcommand{\safenum}[2]{%
    \StrDel{#1}{ }[\tmpstripped]%
    \IfStrEq{\tmpstripped}{}{--}{%
    \IfStrEq{\tmpstripped}{-}{--}{\num[round-precision=#2]{#1}}}%
}

\newcommand{\safeasym}[3]{%
    \StrDel{#1}{ }[\tmpstripped]%
    \IfStrEq{\tmpstripped}{}{--}{%
    \IfStrEq{\tmpstripped}{-}{--}{%
        \num[round-precision=#3]{#1} $\pm$ \num[round-precision=#3]{#2}%
    }}%
}
\usepackage{xcolor}
\usepackage{siunitx}
\newcommand{\gaia}{\textit{Gaia}}
\newcommand{\feh}{[Fe/H]}

\newcommand{\msun}{$M_{\odot}$}
\newcommand{\rsun}{$R_{\odot}$}

\newcommand{\teff}{$T_\mathrm{eff}$}
\newcommand{\kms}{km\,s$^{-1}$}

\newcommand{\gstarDESIshort}{GALEX J114644.2-132116}
\newcommand{\logg}{$\log g$}

\newcommand{\synthe}{\textsc{Synthe}}

\newcommand{\ismissing}[1]{%
  \StrDel{#1}{ }[\tmpstripped]%
  \IfStrEq{\tmpstripped}{}{true}{\IfStrEq{\tmpstripped}{-}{true}{false}{}}%
}
\newcommand{\asymnum}[4]{%
  \StrDel{#1}{ }[\tmpval]%
  \StrDel{#2}{ }[\tmperrp]%
  \StrDel{#3}{ }[\tmperrm]%
  \IfStrEq{\tmpval}{}{--}{%
  \IfStrEq{\tmpval}{-}{--}{%
    \IfStrEq{\tmperrp}{}{%
      $\num[round-precision={#4}]{#1}$%
    }{%
      $\num[round-precision={#4}]{#1}^{+\num[round-precision={#4}]{#2}}_{-\num[round-precision={#4}]{#3}}$%
    }%
  }}%
}

\newcommand{\asymnumtwo}[3]{%
  \StrDel{#1}{ }[\tmpval]%
  \StrDel{#2}{ }[\tmperrp]%
  \IfStrEq{\tmpval}{}{--}{%
  \IfStrEq{\tmpval}{-}{--}{%
    \IfStrEq{\tmperrp}{}{%
      $\num[round-precision=0]{#1}$%
    }{%
      $\num[round-precision=0]{#1}^{+\num[round-precision=0]{#2}}_{-\num[round-precision=0]{#3}}$%
    }%
  }}%
}

\usepackage{xfp}

\newcommand{\safeasymscaled}[4]{%
    \StrDel{#1}{ }[\tmpstripped]%
    \IfStrEq{\tmpstripped}{}{--}{%
    \IfStrEq{\tmpstripped}{-}{--}{%
        \num[round-precision=#3]{\fpeval{#1*#4}} $\pm$ \num[round-precision=#3]{\fpeval{#2*#4}}%
    }}%
}
\usepackage{longtable}
\usepackage{placeins}
\begin{document}

   \title{Intermediate-mass runaway and hypervelocity star candidates in DESI DR1}
   \author{A.Bhat
          \inst{1}\thanks{aakashbhat7@gmail.com},
          M.Dorsch
          \inst{1},
          U.Heber
          \inst{2},
          S.Geier,
          \inst{1}
          H.Dawson
          \inst{1}
          }
 \institute{Institut für Physik und Astronomie, Universität Potsdam, Haus 28, Karl-Liebknecht-Str.\ 24/25, 14476 Potsdam, Germany\\
              \email{aakashbhat7@gmail.com}
         \and
             Dr.\ Karl\ Remeis-Observatory \& ECAP, Astronomical Institute, Friedrich-Alexander University Erlangen-Nuremberg, Sternwartstr.~7, 96049 Bamberg, Germany
             }

 
  \abstract
   {Runaway stars ejected from the Galactic disc span a range of velocities, with two extreme regimes of particular interest. Hyper-runaway stars reach velocities approaching the local Galactic escape speed while remaining gravitationally bound to the Galaxy, while hypervelocity stars have speed which exceed the escape speed, leaving them formally unbound. Studying both populations places stringent constraints on star cluster formation, supernova explosions in massive binaries, and the shape of the stellar halo. Here we present a spectroscopic search for such extreme-velocity stars using data from the first data release of the Dark Energy Spectroscopic Instrument (DESI DR1). Because these objects are typically located at distances beyond the range where \gaia\ parallaxes are reliable, we derive stellar parameters through a combined analysis of spectroscopy and spectral energy distributions (SEDs). These parameters are then compared with stellar evolutionary tracks to infer distances, which are subsequently used to compute the kinematics of the candidates. We identify a sub-sample of eight stars that appear unbound to the Galaxy, most of which are likely of extragalactic origin. These stars exhibit projected rotational velocities exceeding $50$~\kms, consistent with a main-sequence nature. An additional eleven stars are identified as potential hypervelocity candidates under the assumption that they are main-sequence stars. This interpretation is not supported by their rotational velocities, making their unbound status less secure. Finally, $26$ stars show evidence of being main-sequence runaway stars ejected from the Galactic disc, some of which may be hyper-runaway candidates. The fastest object in the sample is a metal-poor hypervelocity star candidate with a Galactocentric velocity upper-limit of $934 \pm 110$~\kms\ at a distance of $95 \pm 13$~kpc. The star with the largest measured radial velocity of $513$~\kms\ is most likely a slowly pulsating B-type runaway star that was ejected from the inner disc $\sim45$ Myr ago.}
   	

\keywords{ Stars: late-type --
                 stars: Population II --
                 (stars:) blue stragglers
               }

\maketitle
\nolinenumbers
\section{Introduction}

The Galactic halo has no ongoing star formation and the last halo formation events are thought to be a few Gyr old. Young and massive O/B-type stars discovered far away from the Galactic disc with high velocities were therefore described as "runaway stars" by \citet{1961BAN....15..265B}. They have historically been defined and selected with space velocities (post solar-motion correction) above 30 \kms and high Galactic latitudes placing them in the stellar halo. The velocity limit is a few times higher than the dispersion velocity of stars within open clusters, where most of them originally form. As the velocity cut increases, the number of observed runaway stars drops significantly. This can be explained by the mechanism through which they are created.

Theoretical velocity limits for runaway stars are of the order $300-500$ \kms, depending highly on how the star was ejected. Formation channels include dynamical ejections (DES) in proto-stellar clusters \citep{poveda1967}, where dynamical binary-binary and binary-single star interactions lead to the ejection or ejections of one or more stars. A second channel of formation is through the binary supernova scenario (BSS) when the more massive primary explodes in a Type II supernova \citep{1961BAN....15..265B}. A main sequence runaway can, however, also be formed in a similar process as a donor to a white dwarf companion which undergoes a Type Ia explosion \citep{2000A&A...362.1046L}. There is also a less well-studied double-ejection process where the primary of a dynamically-ejected binary undergoes a supernova explosion ejecting the secondary \citep{2010MNRAS.404.1564P}. 

Dynamical ejection from clusters has been studied combining observation with kinematic computations \citep{hoogerwerf2001,Bhat2022}. Similarly, binary supernova ejections have also been confirmed \citep{Dincel2015,Neuh2020,schoettler2020,Farias2020}. Studies of double ejections are so far sparse \citep{Dorigo2020}, owing to the difficulty in reconstructing the trajectory due to the two separate acceleration events. Dynamical ejections typically occur during the early stages of a cluster, whereas binary supernova ejections require the primary to evolve to the supernova stage, typically taking $\sim 10$--$40$~Myr. Consequently, the flight time of a dynamically ejected star is expected to be comparable to its age.

In the last decade, studies have found OB-type runaways close to or even faster than the theoretical speed limits for the proposed scenarios \citep{2008ApJ...684L.103P,2018AJ....156...87L,2019ApJ...873..116H,irrgang19,2021A&A...646L...4I,Raddi2021}. These stars, also called hyper-runaway stars, are very rare, and only a few tens of them are known \citep{2020A&A...637A..53K}. Initially the term hyper-runaway was used to refer to stars close to or exceeding the Galactic escape velocity \citep[$497\pm8$ \kms\, around the solar circle as determined by ][]{2021A&A...649A.136K} which were formed in the Milky way disc. This was to separate them from hypervelocity stars (HVSs). Originally, HVS stars referred to extremely fast stars produced through the tidal disruption of a binary by the supermassive black hole at the centre of the Galaxy in the Hills mechanism \citep{Hills,2015ARA&A..53...15B}. The first candidate was discovered by \citet{2005ApJ...622L..33B}, triggering the HVS survey \citep{2006ApJ...647..303B}. Subsequent studies uncovered many more candidates, but only one has been conclusively confirmed to be consistent with a Galactic centre origin \citep{2020MNRAS.491.2465K}, while for the majority a Galactic centre ejection has been shown to be inconsistent \citep{2020A&A...637A..53K}. More recently, HVS candidates produced through the Hills mechanism were found by \citet{2026A&A...709A.117C} and \citet{2026ApJ..1003L...9D} by combining \gaia\, data with spectral data from the Dark Energy Spectroscopic Instrument (DESI) \citep{DESI2025}. 

The term hypervelocity star, however, has been used for all stars unbound to the Galaxy, irrespective of their place of origin \citep{2005ApJ...634L.181E,2005A&A...444L..61H}, and even their evolutionary state \citep{2026OJAp....964326E}. Here, we continue with this notation of HVS. For some stars, the Hill mechanism might still be in play. In particular, it was recently shown that many of the HVS candidates could instead have been ejected from the Large Magellanic Cloud \citep[LMC][]{2025ApJ...982..188H,2025ApJ...993L..10L} through the Hills mechanism, suggesting the presence of an SMBH at the centre of the LMC. This is consistent with previous studies, such as that of HVS 3 \citep{2005ApJ...634L.181E, 2008A&A...480L..37P,2018A&A...620A..48I} and simulations of HVS ejections from the LMC and the Milky Way \citep{2021MNRAS.507.4997E}. A similar behaviour has been seen in a sample of hypervelocity RR Lyrae stars discovered recently by \citet{2025ApJ...994..148F}.  

In lieu of this, we use the term hyper-runaway star here to refer to any main-sequence (MS) star close to but not exceeding the Galactic escape velocity which was ejected from the disc, similar to what was done by \citet{2023MNRAS.518.6223I}. The exact mechanism of their ejection is not known. Studies show that binary parameters (like common-envelope efficiency) must be different than those presently supported in the literature, to produce these objects through the BSS \citep{2020MNRAS.497.5344E}. Similarly, dynamical cluster ejections can produce them, but not at the rate observed so far \citep{2012ApJ...751..133P}.

The most reliable candidates have historically come from radial velocity (RV) surveys preceding Gaia, with pre-Gaia studies deriving proper motions from a combination of multiple astrometric surveys \citep{2011MNRAS.411.2596S}. Proper-motion-based searches have since proven considerably less robust \citep{2015A&A...576L..14Z}. \citet{2018MNRAS.479.2789B} showed that spurious astrometric solutions in \gaia\, DR2 could mimic hypervelocity star signatures, undermining many candidates selected from that release. More recent studies have mainly relied on the third data release of Gaia \citep[DR3][]{Gaia3} proper motions and distances to shortlist candidates, followed by spectral follow-up for the fastest candidates and, in some cases, using the RV from Gaia for cooler MS stars \citep{2021A&A...646L...4I,marchetti2021,Marchetti2022,2023A&A...679A.109C}. However, \gaia\, distances remain unreliable for a substantial fraction of these targets, limiting the robustness of DR3-based selection on its own. Even though \gaia\, DR4 parallaxes are expected to be $\sim 1.33$ times more precise than DR3 \citep{2022MNRAS.512.2350E}, the distances at which most HVS candidates are found will still remain beyond what \gaia\, can reliably probe. In a few cases, candidates have been directly found using spectral data, for example with the Large Sky Area Multi-Object Fiber Spectroscopic Telescope (LAMOST) \citep{2024ApJS..272...45G,2025ApJ...986...22S}.

In this paper, we leverage only the colour data from \gaia\, DR3 and spectra from the first data release (DR1) of DESI to search for hyper-runaway stars. We model the spectra of all the stars and combine that with spectral energy distribution (SED) modelling and evolutionary tracks to deselect halo contaminants like blue horizontal branch (BHB) and other late type stars. Where possible, we use data from the Zwicky Transient Facility \citep[ZTF;][]{2019PASP..131a8002B} to search for light-curve variability.

The paper is structured as follows. In Section 2 we describe how the targets were selected and analyzed. Section 3 summarizes our analysis pipeline. Section 4 describes the results of our spectroscopic and photometric analysis. In Section 5 we utilize evolutionary tracks and describe the most likely states of these stars along with their kinematics. Section 6 focuses on a more in-depth analysis of individual stars which are particularly fast. In Section 7 we summarize our results and provide a general outlook to the field of hyper-runaway stars.

\section{Sample Selection}

We used the \gaia\, database to search for stars which have been observed with DESI in their DR1. Our aim is to identify young stars in the Galactic halo. These stars have historically, along with massive O-type stars in the disc contributed the most to the total number of runaway stars known. Since stars from early O- to mid A-types live at most up to a $\sim500$ Myr, any such star in the halo or moving away from the disc with a high velocity must be a runaway star. The halo, however, also harbours older, evolved stars like blue horizontal branch (BHB) stars \citep{2019MNRAS.490.5757S,2024A&A...685A.134C} as well as metal-poor main-sequence and giant stars \citep{2015MNRAS.447.2046H,2018MNRAS.481.1028H}. Many of these stars can originate from accreted galaxy mergers \citep{2018Natur.563...85H}, boosting their velocities and causing them to appear as contaminants in our samples. These have not been easy to disentangle from the normal halo population. Metallicity estimates could help, but these estimates require high S/N, high-resolution spectra, which are not available for most candidates. Therefore, to minimise contamination from fast halo stars, we restrict our sample to objects with $G_{\rm BP}-G_{\rm RP} < 0.1$, corresponding to effective temperatures $T_{\rm eff} > 9500$~K \citep{2018A&A...616A...8A}. Stars cooler than this limit can be sufficiently old to have either formed in situ in the halo or been accreted, and are therefore excluded. This selection yields $\sim27,000$ targets after cross-matching \gaia\ DR3 and DESI-DR1.

DESI is a multi-object spectrograph mounted on the Mayall 4-meter telescope at Kitt Peak National Observatory. It is capable of simultaneously measuring spectra for up to 5,000 targets using robotically positioned fibers. Even though it was designed primarily to constrain dark energy via observations of millions of galaxies, DESI is also obtaining millions of stellar spectra as part of its Milky Way Survey. The DESI spectra have an average resolving power of $R \simeq 2000$ at 300~nm and $R \simeq 5500$ at 980~nm, making them well suited for deriving radial velocities with a typical precision of $< 10$ \kms. When combined with \gaia\ positions, parallaxes, and proper motions, this provides the full six-dimensional phase-space information required for a kinematic analysis. Because many halo objects are distant and thus lack reliable parallaxes, spectrophotometric distances were derived by jointly modelling the spectra and SEDs, and utilizing stellar evolutionary tracks to estimate stellar masses. For this purpose, we retained only spectra with a signal-to-noise ratio $\mathrm{S/N} > 10$, which were then analysed using an automated stellar model-fitting pipeline.

We use four different pre-computed stellar model grids for spectral modelling by fitting synthetic spectra to the observed spectra as
described in \citet{Irrg2014}.

\begin{itemize}
\item[1] Following-up on the detailed analyses of run-away B type MS
stars \citep{irrgang19,irrgang21,Raddi2021}, we used the
the hybrid non-LTE grids of \citet{2018A&A...615L...5I} and
\citet{2020A&A...633L...5I} computed with the
\textsc{Atlas/Detail/Surface} (ADS) hybrid LTE and non-LTE stellar
atmosphere and spectral synthesis codes \citep[][and references
therein]{irrgang21}.
These grids span $10,000$ to $33,000$ K in \teff\, and include the most important
metals (C, N, O, Ne, Mg, Al, Si, S, Ar, and Fe) synthesized in NLTE.
This allows for metal abundance determinations by fitting the
metallicity \feh\ in the range from $-1.0$ to $+0.5$ dex).

\item[2] A second set of models extends below 9,000 K ( $4,600-14,000$ K and \logg\, between $2.8-6.0$ dex), only accounting for LTE by means of the
\textsc{Atlas12} and \synthe\ software (Kurucz 1993) including an extensive list of metal lines \citep[see][]{dorsch21}. Recent improvements for the computation of level dissolution in the proximity
of the Balmer jump introduced by \citet{2018A&A...615L...5I} were taken into account.
This \synthe\ grid samples solar-scaled metallicity between \feh$=-2.0$
and \feh$=0.5$ dex.

\item[3]
The third grid is the so-called ``2nd generation Bamberg model'' grid \citep{heber26b}, also computed with the \textsc{Atlas/Detail/Surface} (ADS) hybrid LTE/non-LTE stellar atmosphere and spectral synthesis codes. It has been used to analyse large samples of hot subdwarf stars
\citep{dawson26,latour26,heber26b}. Spectrum synthesis was limited to the H- and He-lines, allowing for a very large range of He-abundances ($\log He/H = -5.0$ to $+2.0$) in addition to large ranges of \teff\, and \logg\, \citep[see Fig. 4 of][for grid
dimensions]{heber26b}, covering $9,000$\,K to $105,000$\,K and \logg\, up to $7$ dex. This grid is useful to identify He-peculiarities known to occur in hot subdwarfs, BHB stars, and ELM WDs.

\item[4] For those stars we identified as white dwarfs from their spectra, we used the grids computed for DA and DO white dwarfs, described in \citet{2016A&A...587A.101R}.

\end{itemize}

Projected rotation velocities are important to identify the nature of stars. They are derived from H, He, and metal lines in grids 1 and 2. Results from grid 3 are less precise because they are derived from H and He
lines only. To account for systematic uncertainties of the resulting atmospheric parameters we adopted the estimates from \citet{dawson26} and added them in quadrature to the statistical uncertainties.

After removing cosmics and data reduction artifacts, the entire spectrum was fit using the Interactive Spectral Interpretation System \citep{2000ASPC..216..591H}. We use a $\chi^2$- minimization routine, similar to that described in \citet{Irrg2014}. We simultaneously fit stellar parameters \logg, \teff, $v \sin i$, $v_{\rm rad}$, and in some cases the helium abundance. \feh\, is fit with the B-type and Synthe grids. For the resolution of the DESI spectra, $v \sin i$ below $\sim50$ \kms\, can not be confidently detected. 

Based on the spectroscopic parameters we removed high-gravity white dwarfs and hot subdwarfs and only selected objects with radial velocity RV$>150$ \kms. We then construct two sub-samples. The first sample of runaway star candidates is selected by requiring \feh $>-0.3$, ensuring a disc-like metallicity consistent with a runaway origin. The second sample consists of halo HVS candidates, which is assembled by individually inspecting stars based on their derived distances and kinematics, and therefore includes stars with \feh $<-0.3$. The initial union of these two samples left us with 197 stars. We then cross-matched those with the SIMBAD database to check for known stars. We filter out all known systems where either a spectral classification was already known or a radial velocity was provided. This allowed us to filter out 45 stars among which were known high velocity stars, extremely low mass white dwarfs \citep[ELM WDs][]{Brown2010ELM}, eclipsing binaries, and horizontal branch (HB) stars. The SIMBAD results were validated individually. The results for the full DESI sample will be detailed in a forthcoming paper. Here we report only on the sub-sample expected to contain runaway stars not known before rather than halo stars of extreme velocity. We note here that the DESI broad-line issue identified by \citet{2026ApJ..1000..216K} appears specifically for hot DA white dwarfs. The problem exists in broad hydrogen line profiles, vanishing for cooler, narrower-lined white dwarfs \citet{2026ApJ..1006..248K}. Thus, we do not expect this issue here.

\begin{figure}
   \centering
   \includegraphics[width=0.48\textwidth]{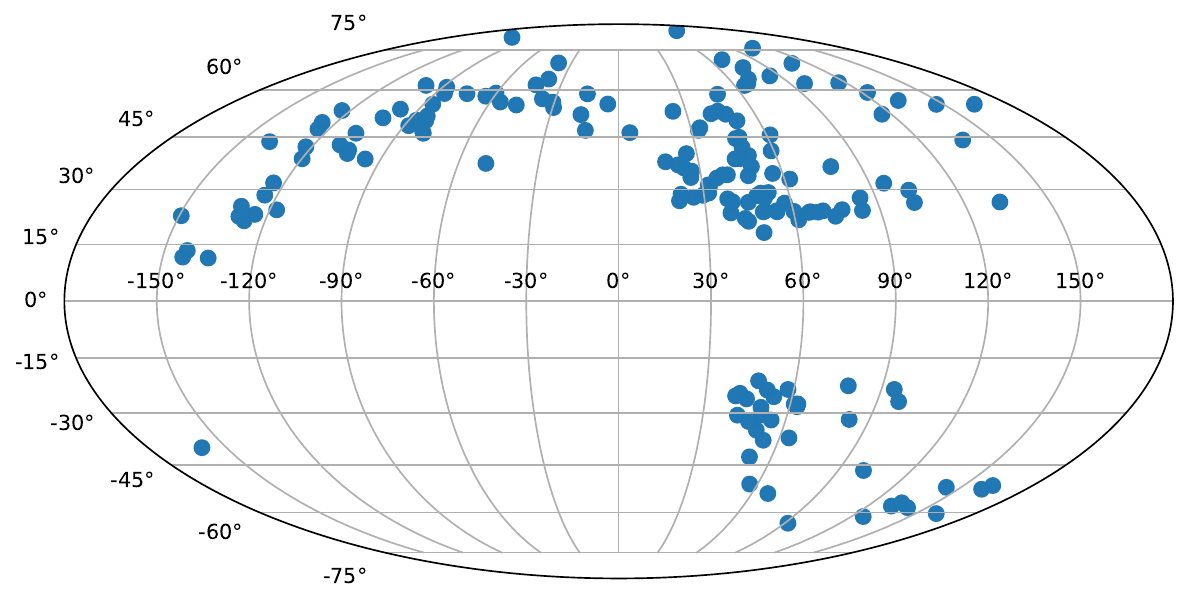}
   \caption{Sky distribution of the selected sample in galactic coordinates. The sample is overwhelmingly outside the disc due to the DESI footprint.}
    \label{fig:lb}%
    \end{figure}

\section{Follow-up analysis of the sample}
\label{sec:three}
Our analysis of the sample consisted of a four-step procedure. The spectral analysis described above was combined with photometry, and mass and radius estimates from evolutionary tracks, and kinematic analyses.

\subsection{Spectral energy distribution analysis}
We used the SED fitting procedure described in \citet{Uli2018} for all the 152 stars to filter out contaminants. The photometric fitting used the same model grids as the spectral fitting routine. SED fits were constrained using the atmospheric parameters derived from the spectroscopic fits along with their corresponding systematic uncertainties. The angular diameter $\log \Theta$ and the reddening E(B-V) were kept as free parameters. We used the extinction law of \citet{2019ApJ...886..108F} for a reddening parameter of $R_{55} = 3.02$, to account for interstellar extinction. For cases where a reliable \gaia\, parallax with less than $20\%$ uncertainty was available, we were able to filter out BHB and late-type stars using the spectro-photometric masses derived using the radius $R=\frac{d\Theta}{2}$ and the $\log g$, as $M=\frac{gR^2}{G}$. We identified 35 BHB candidate stars. Since main-sequence stars can have cooler companions which may not be visible in the spectra but may contribute in the infrared, the SED fits also consider the flux contribution of a companion star. These were modelled using the Phoenix spectral library \citep{2013A&A...553A...6H}. These binary grids were used to filter out any composite binaries if a binary fit was preferred. We also flagged objects with a colour excess which might represent contamination and objects with a lack of photometric data. Stars with a colour excess significantly higher than the reddening maps provided by \citet{2011ApJ...737..103S} were filtered out, leaving a sample of 54 stars.

\subsection{Comparison with stellar evolution tracks}

\begin{figure}
   \centering
    \includegraphics[width=0.47\textwidth]{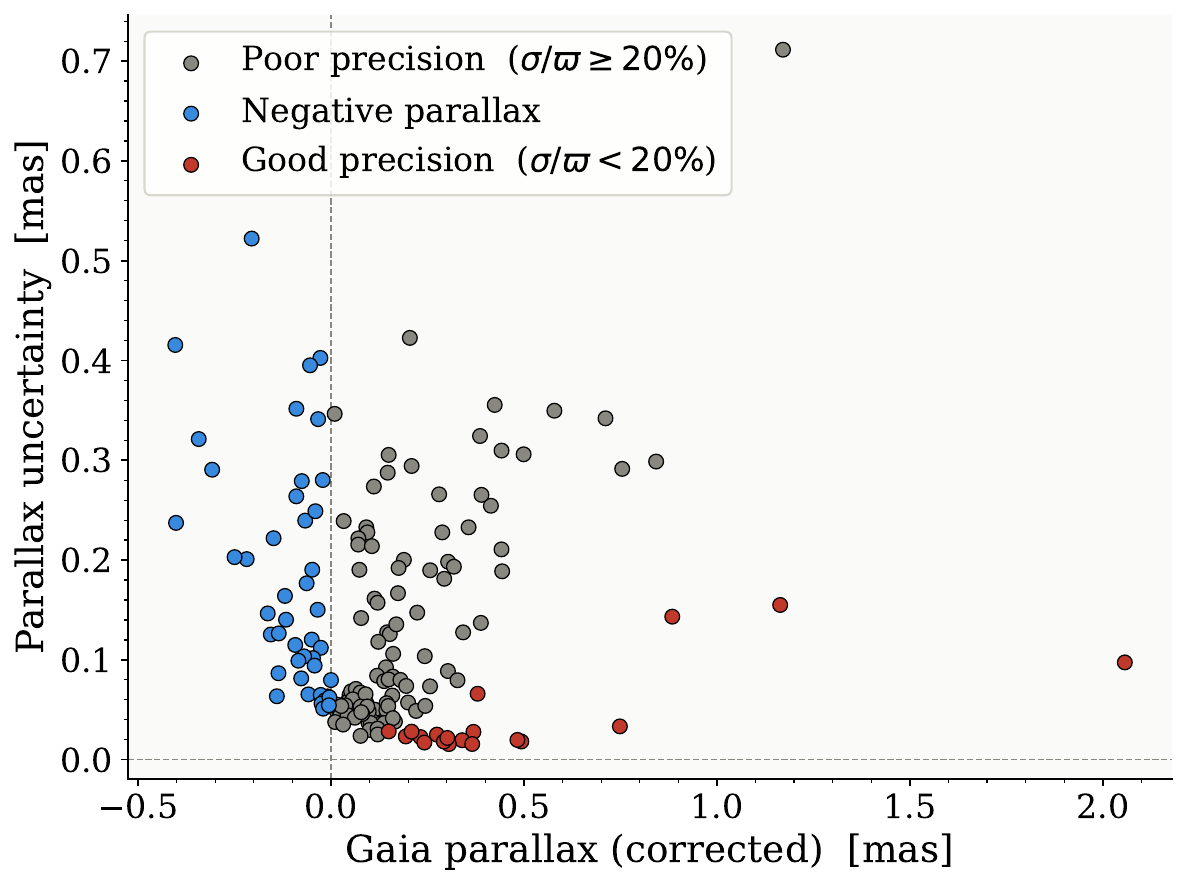}
   \caption{\gaia\, parallax uncertainties for our sample as a function of the \gaia\, parallax. These parallaxes are zero-point corrected according to \citet{Lindgren2021} using the corresponding python package. }
              \label{fig:par}%
    \end{figure}

\gaia\, parallaxes are shown in Fig.~\ref{fig:par}. Only $7$ stars have parallaxes with less than $20\%$ uncertainty. Many stars have negative parallaxes which are difficult to sample without a high amount of uncertainty and are biased towards high distances. To determine accurate spectro-photometric distances we first derived estimates of masses and ages for all $54$ stars based on MESA Isochrones and Stellar Tracks
(MIST) evolutionary tracks \citep{MIST0,MIST1,mesa1,mesa2,mesa3}. A caveat here is that MIST tracks assume that no interaction has happened before which substantially changed the structure of the runaway. Mass transfer from the primary could indeed change the parameters of the runaway but the effects should only hold for a thermal time before the star behaves as a normal MS star again. This has been shown for case studies of binary systems, for example by \citet{2024A&A...687A.222W}. Evolutionary tracks with solar metallicity are fit utilizing a Bayesian MCMC approach similar to what was done in \citet{2025A&A...700L..23B}. We interpolate to sample near the log $g$ and T$_{\rm eff}$ we derive from the spectra. We use the SciPy package \citep{2020SciPy-NMeth} along with the Emcee package \citep{emcee} for this purpose. Combining the mass with the angular diameter, we derived spectro-photometric distances for these stars, given as $D = 2R/ \Theta$, where $\Theta$ is the angular diameter on the sky.
\subsection{Kinematics}

For stars where the distance could be derived (47 stars) or a reasonably accurate parallax ($<20\%$ uncertainty) was available (7 stars), we performed a kinematic analysis utilizing the Milky way Model I of \citet{andreas} in galpy \citep{2015ApJS..216...29B}\footnote{Galpy approximates the halo potential by replacing the strict cut-off at $200$ kpc with a potential that decreases to infinity. This does not affect our stars because within $100$ kpc, the change in escape velocity is less than $0.3$\%.}. Model 1 of \citet{andreas} is a revised version of the mass model of \citet{1991RMxAA..22..255A}. The local standard of rest (LSR) velocities (U,V,W)$_\odot$ are $(11.1\pm1, 12.24\pm2,7.25\pm0.5)$ \kms\,taken from \citet{2010MNRAS.403.1829S}. We also compare the results in each case with those derived using the Mcmillan17 models \citep{2017MNRAS.465...76M}. The differences are generally small, with the main difference being that Model I of \citet{andreas} has a higher escape velocity than Mcmillan17. This impacts the probability of the star being bound to the Galaxy, but does not have a significant difference on ejection velocities or time of flight. As such, Model I of \citet{andreas} serves as an upper bound, and if a star is unbound to the Galaxy in this model, then it will be unbound in \citet{2017MNRAS.465...76M} as well. We therefore stick to Model 1 of this paper. For the cases where the distance uncertainty was less than $10\%$, we searched for open clusters that might have been the sites of origin of these stars, according to the procedure described in \citet{Bhat2022}.

\section{Spectro-photometric results}

\begin{figure}
\centering
    \includegraphics[width=\linewidth]{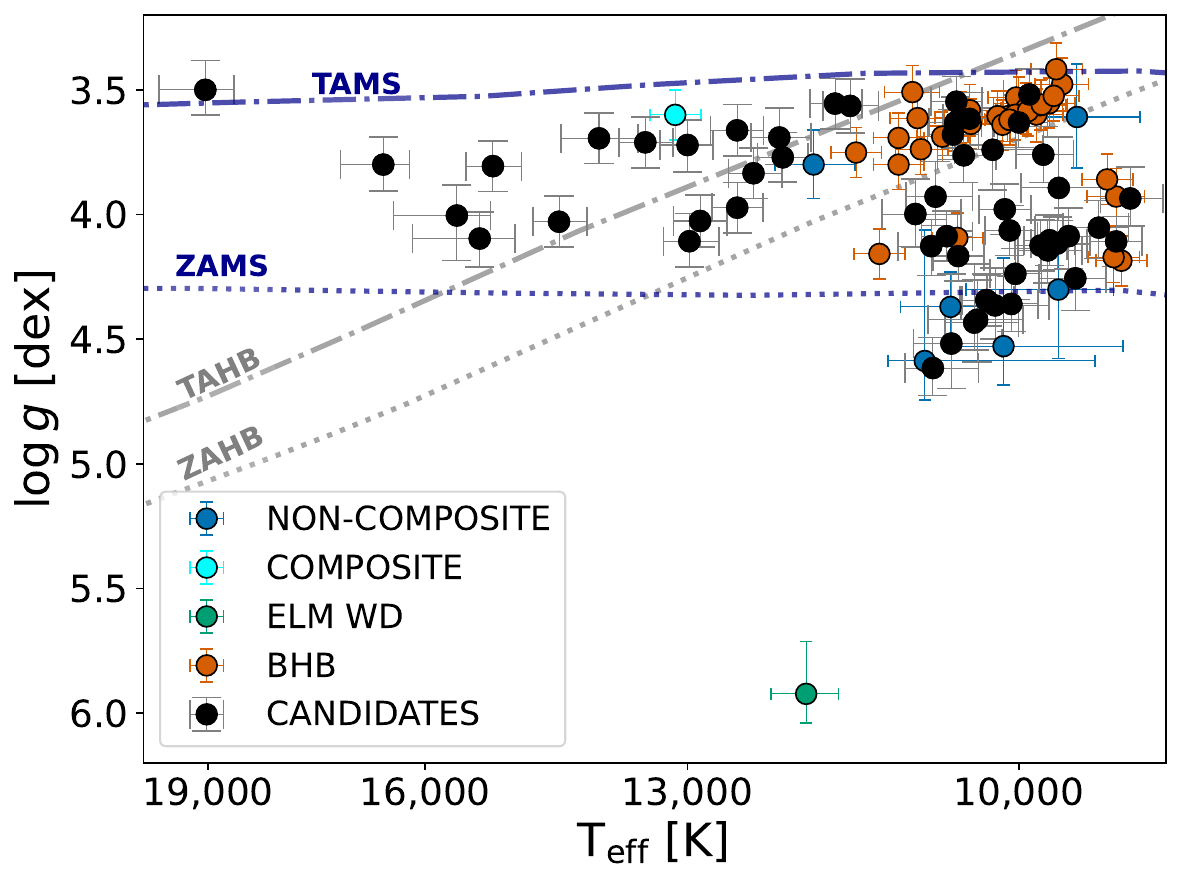}
    \includegraphics[width=\linewidth]{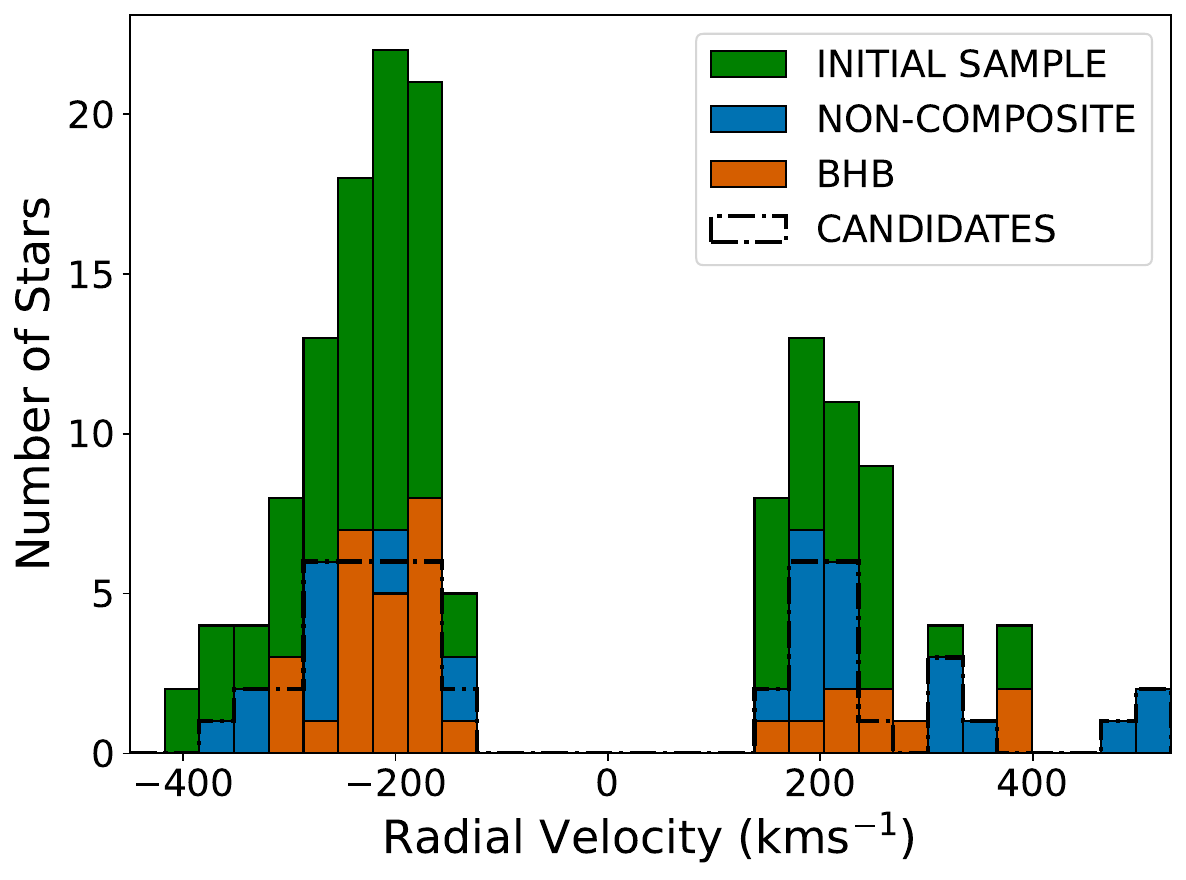}
 \caption{Upper panel: Kiel diagram of our selected sample. BHB (orange) and ELM WD (green) stars are marked according to masses calculated through parallaxes. The hot runaway candidates are plotted in black. Lower panel: Distribution of barycentric radial velocities for the sample. Our sample has a slight excess of negative radial velocities, which could be due to the halo lagging behind Galactic rotation in the disc. }
  \label{fig:kielRV}
\end{figure}

Since we analyzed stars which had an $|{\rm RV}|$ greater than 150 \kms\, this meant that either the stars belong to the old halo population, or that they are younger stars in the halo ejected out of the disc. The spectro-photometric analysis allows us to distinguish between those populations. The Kiel diagram of our sample is shown in Fig.\ref{fig:kielRV}. The zero age main-sequence (ZAMS) and the terminal age main-sequence (TAMS) from the MIST evolutionary tracks are plotted along with the zero age horizontal branch (ZAHB) and terminal age horizontal branch (TAHB) from the evolutionary models of \citet{1993ApJ...419..596D}. The latter are plotted for a helium mass fraction of $Y=0.247$ and solar metallicity. Due to our colour-excess filtering, all except one of the stars show single, non-composite SEDs. Three ELM WDs were found in this sample, two of them already known
 (and are not plotted). An example fit of the well-studied ELM WD SBSS 1434+503 \citep[studied by][]{2007ApJ...660.1451K} is shown in Fig.\ref{fig:examplefit}. ELM WD are found in close binaries \citep{2020ApJ...889...49B} and are often also fast halo stars, but they being RV-variable their system velocities need to be known to determine their kinematics. Most of the BHBs were within the ZAHB and TAHB, but seven of them lie below the ZAHB. The corresponding barycentric RVs are plotted in the lower panel of Fig.~\ref{fig:kielRV}. An asymmetry in the RVs is seen with more objects having negative RVs, implying that they are coming towards us and are most likely bound halo objects. Our final sample of candidate runaway stars (S/N $>10$) is shown as black dots in Fig.\ref{fig:kielRV}.

\section{Evolutionary status}

\begin{figure}
\centering
    \includegraphics[width=0.49\textwidth]{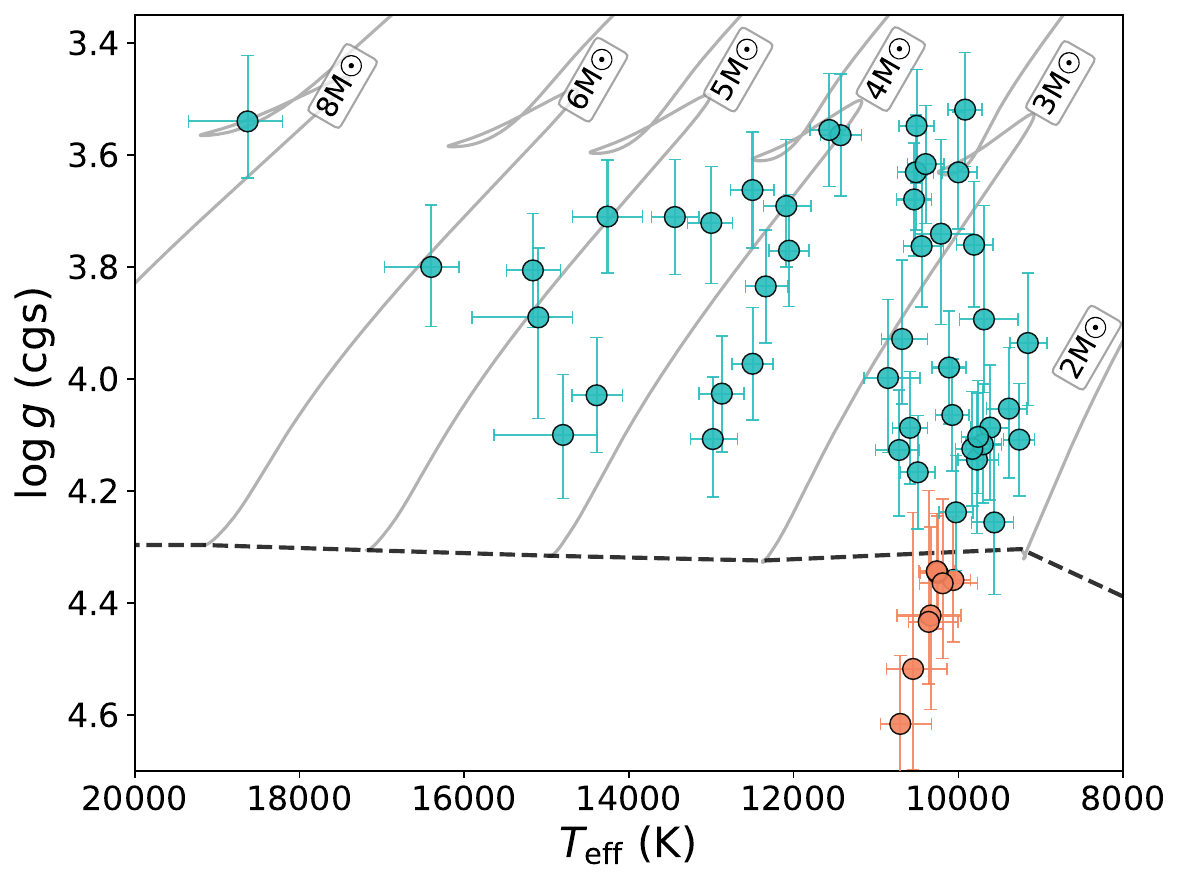}
    \caption{Kiel diagram for the selected stars in the sample with the ZAMS plotted as dashed line and the solar metallicity MIST tracks for different masses plotted as solid lines. Coral points mark those below the ZAMS which are not considered further. }
    \label{fig:kielwithtracks}
\end{figure}

Fig.~\ref{fig:kielwithtracks} shows the Kiel diagram of the sample with evolutionary tracks overlaid. These tracks were used for sampling the posteriors of the masses, ages, and radii of the stars by comparing the predicted $\log g$ and \teff\, with the observed values. The dashed black line which marks the zero-age main-sequence (ZAMS) and shows that 8 of our stars are most likely not main sequence stars but could belong to a pre-ELM WD population moving towards the cooling track. We do not consider these stars any more. 

\begin{figure}
    \centering
    \includegraphics[width=0.49\textwidth]{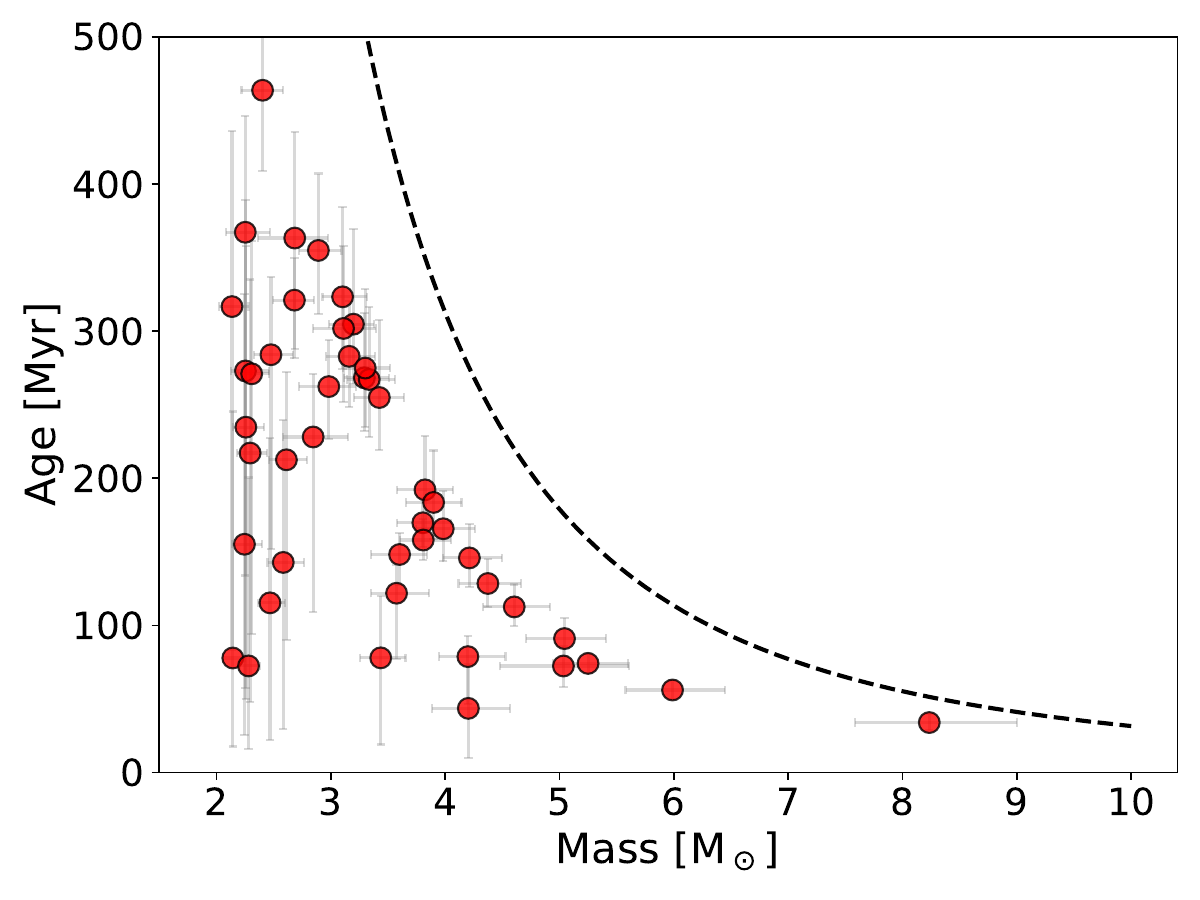}
    \caption{Age estimates of the sample as a function of mass. The high uncertainty for stars close to $2$~\msun\,arises from these stars being close to the ZAMS. The black dashed curve is an approximate MS lifetime power law $t_{\rm MS}=10^4(M/M_\odot)^{-2.5}$ Myr.}
    \label{fig:mass}
\end{figure}

The masses and computed ages for all the stars are shown in Fig.~\ref{fig:mass}. All of the stars in our sample are more massive than $2$ \msun\,. Stars more massive than $3$ \msun\, are close to the terminal age main-sequence (TAMS). This is possibly a selection effect since these stars need some time to reach higher galactic latitudes. There are 3 stars between 2 and 3 \msun\, which are still near the ZAMS. Since these stars can have higher hydrogen burning ages (up to $\sim1$ Gyr for solar metallicity evolutionary tracks) these stars could still qualify as disc runaway candidates. They could also be pre-ELM WDs passing through this part of the evolutionary phase, as was found for one HVS candidate by \citet{2021A&A...650A.102I}. However, more spectra will be needed to characterize such systems.

\begin{figure}
\centering
    \includegraphics[width=\linewidth]{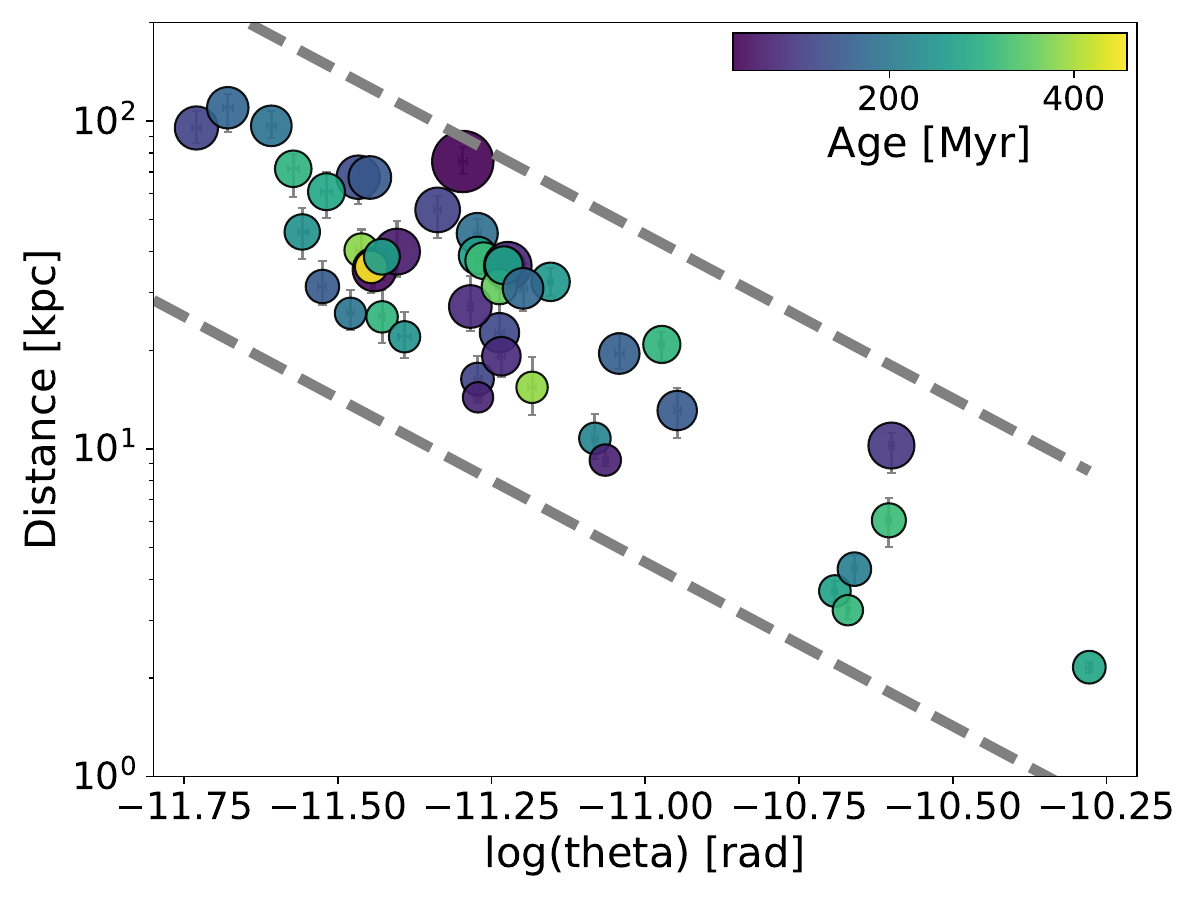}
    \includegraphics[width=\linewidth]{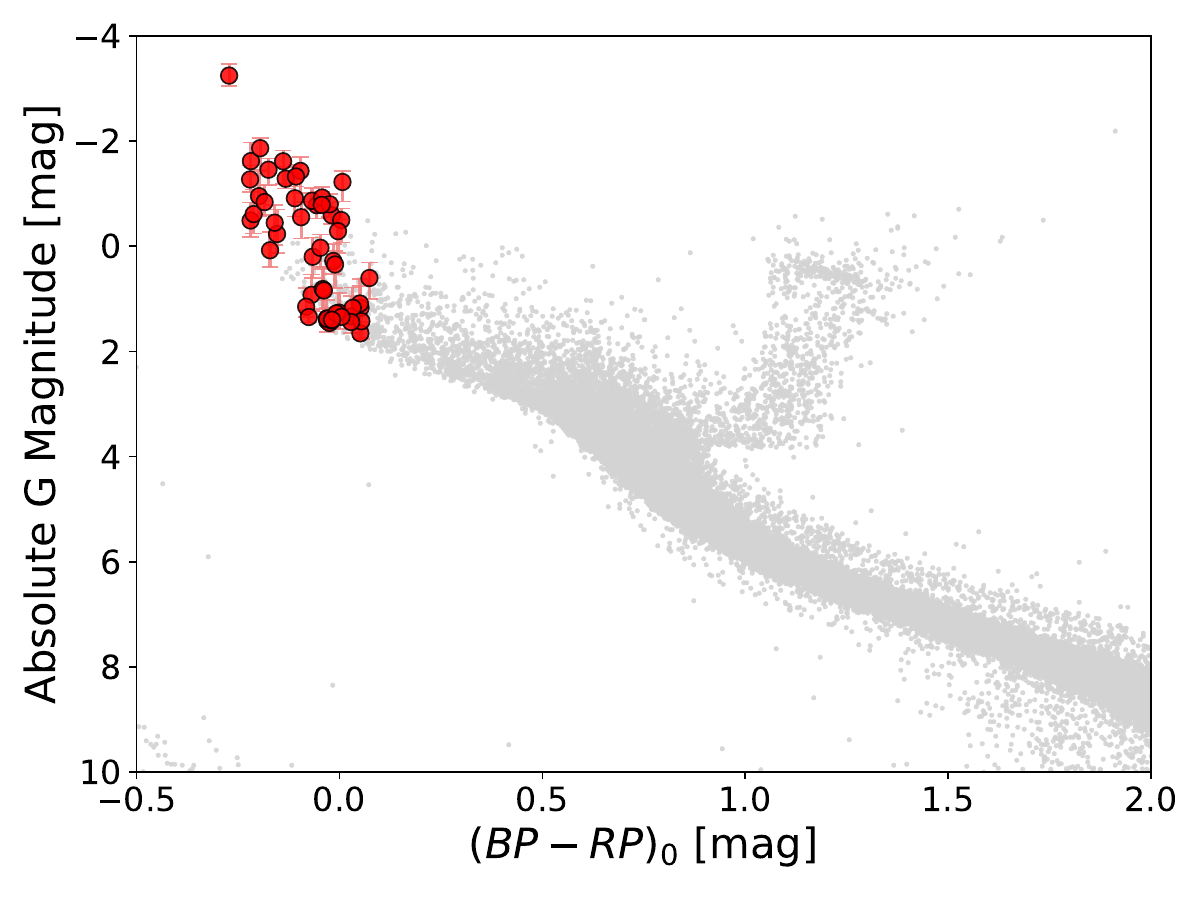}
\caption{Upper panel: The derived distances for our objects as a function of the log of angular diameter. Dashed lines are lines of constant radii of $1$ \rsun\, (lower) and $10$ \rsun\, (upper). Points are coloured by age and their relative sizes reflect the masses. Lower panel: Extinction-corrected \gaia\, CMD with a $100$ pc sample randomly selected in the background (grey points). Uncertainties are mainly due to the uncertainty on distance.}
\label{fig:cmd}
\end{figure}

The spectro-photometric distances computed using the track-derived radii are shown as a function of angular diameter in the upper panel of Fig.~ \ref{fig:cmd}. Most stars in our sample are farther away than $10$ kpc. The corresponding colour-magnitude diagram (CMD) created through the \gaia\, apparent magnitude in the g-band and the $G_{\rm BP}$-$G_{\rm RP}$ colour difference is shown in Fig.~\ref{fig:cmd}. The CMD corroborates the spectral type of our stars.

\section{Kinematics}

\begin{figure}
   \centering
    \includegraphics[width=0.5\textwidth]{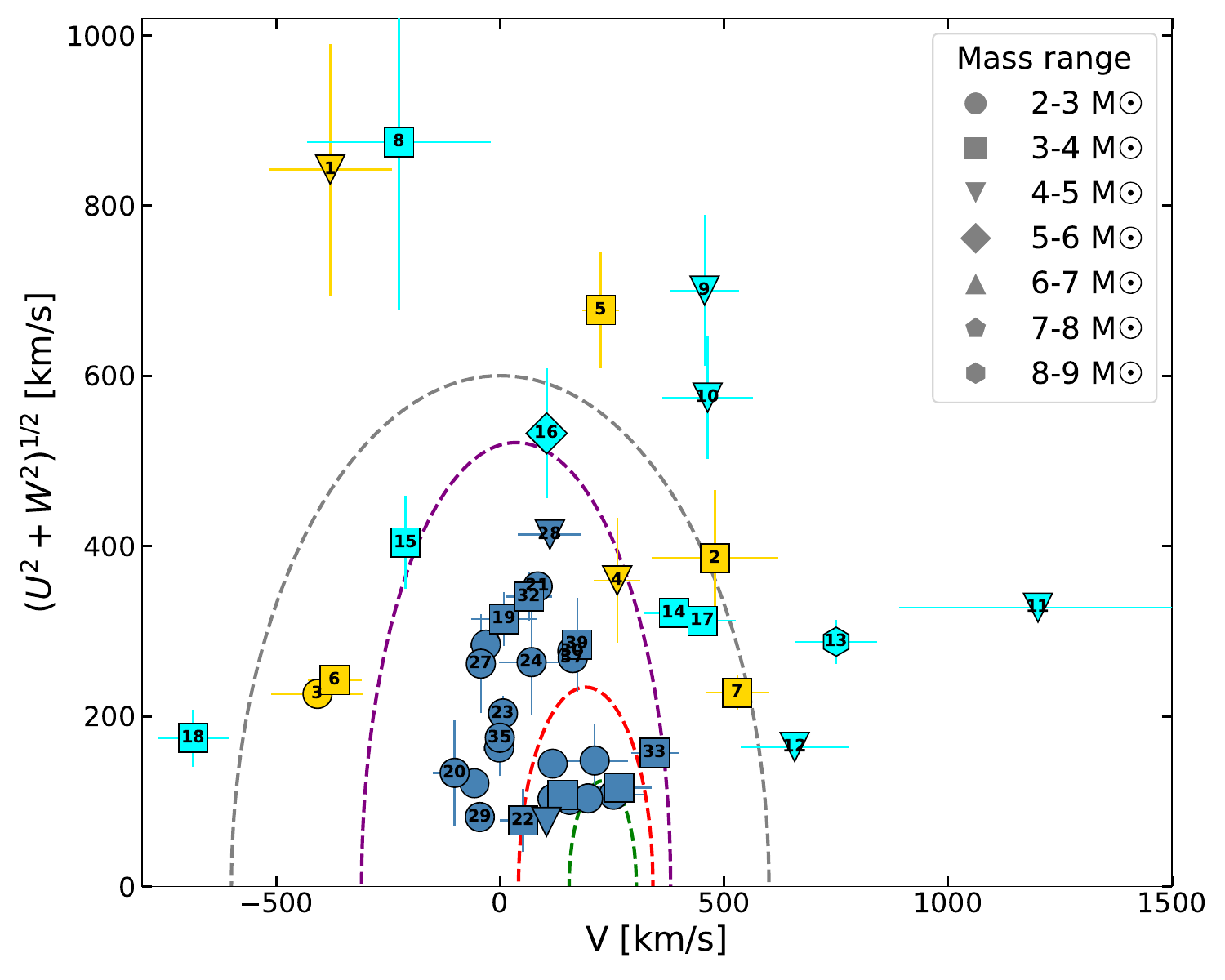}
   \caption{Toomre diagram of the MS runaway candiates (blue), HVS candidates with distances derived spectro-photometrically (gold), and HVS candidates which have $v\sin i<50$ \kms\,(cyan). The contours shown are $3\sigma$ for thin disc (green), thick disc (red), and halo (purple) from \citet{angiano2020}. The grey dashed contour is a Galacto-centric velocity of $600$ \kms. Different markers represent the mass range of the stars. None of the stars lie within the thin disc despite their high metallicities. For clarity only stars outside the thick disc are numbered.}
              \label{fig:toomre}%
    \end{figure}

\begin{figure}
    \includegraphics[width=0.49\textwidth]{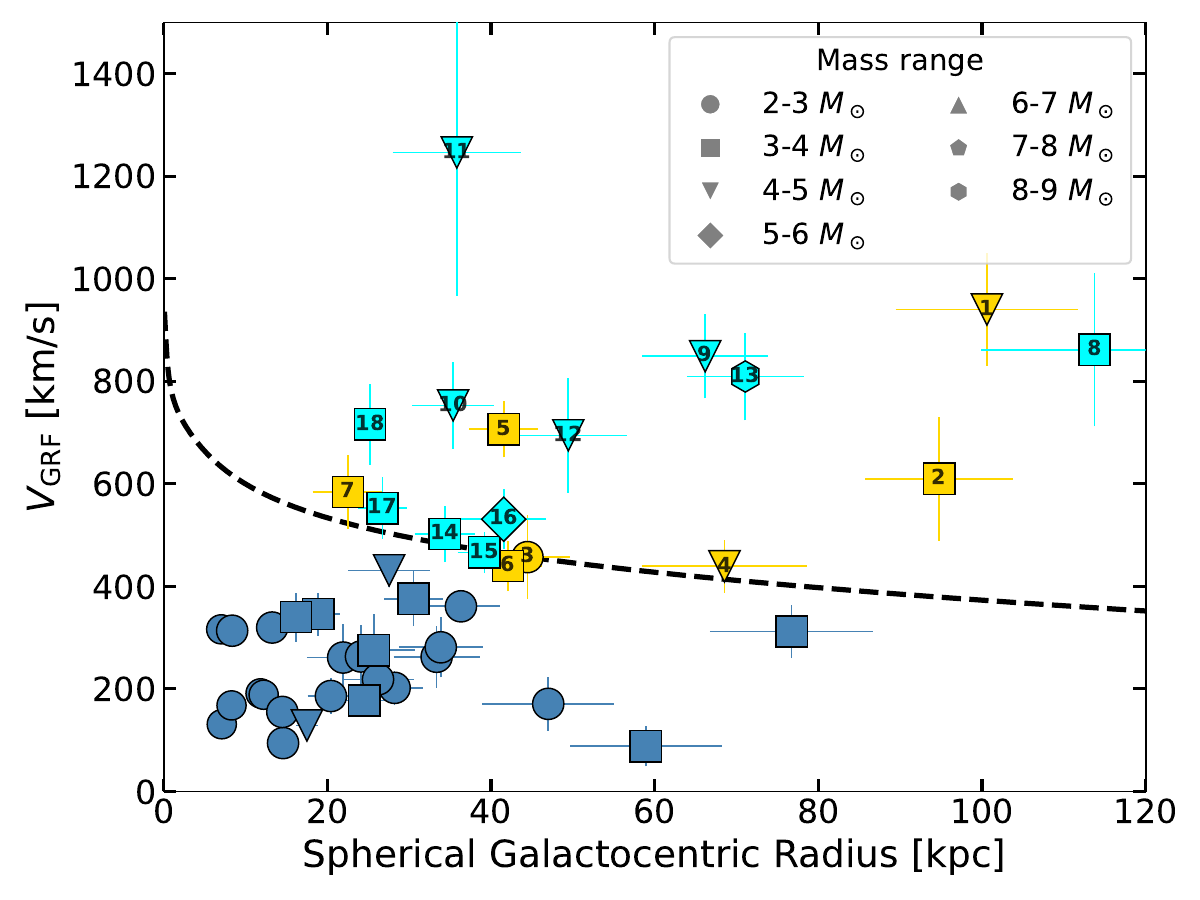}
    \includegraphics[width=0.49\textwidth]{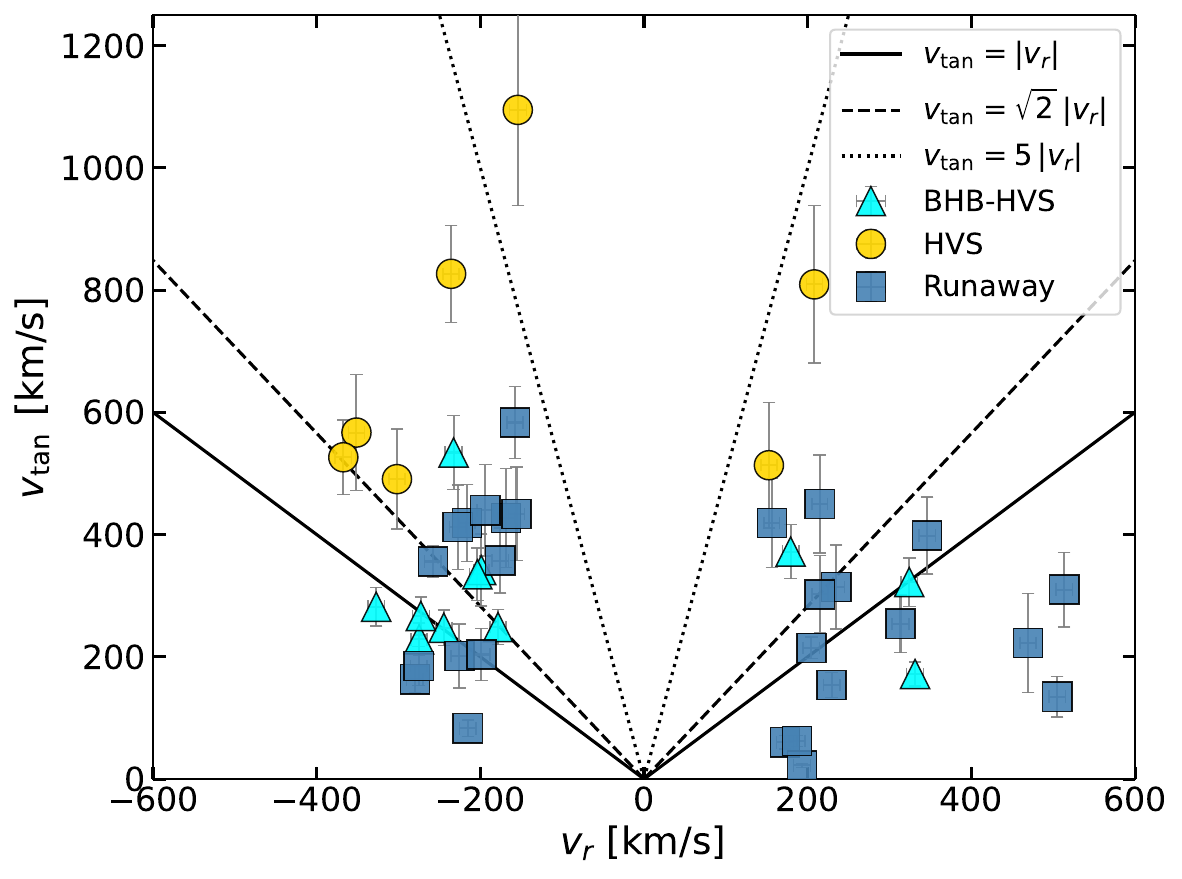}
    \caption{Upper panel: Galacto-centric rest-frame velocities for the sample stars as a function of Galactic radius. Lower panel: Tangential velocity vs. radial velocity but assuming a mass of $0.5$ \msun\, for the BHB-HVS candidates. Solid, dashed, and dotted lines show $v_{\rm tan}$ equal to $1$, $\sqrt{2}$, and $5$ times $v_{\rm rad}$. $v_{\rm tan}=\sqrt{2} v_{\rm rad}$ represents an isotropic distribution.}
    \label{fig:escp}
\end{figure}

The Toomre diagram of the sample is shown in Fig. \ref{fig:toomre}. The dashed curve shows a Galactocentric velocity $v_{\rm grf}>600$ \kms\, and is close to the Galactic escape velocity at the solar circle. Nine stars are outside of this contour and are most likely hypervelocity star candidates. The rest of the sample has halo or thick disc kinematics. Since we use \gaia\, proper motions, the reliability of the solutions is important. The renormalized unit weight error (RUWE) is a quality criterion which is used to quantify deviations from  single-star astrometric solutions. This could be due to binarity but may also be due to measurement errors. The limit for reliability is around $1.4$, as discussed in \citet{gaia2}, although sky-varying cut-offs may be lower \citet{CG2024}. The maximum RUWE value for our sample is 1.07, much lower than any cut-off.
The velocity distribution as a function of radius is shown in Fig.~\ref{fig:escp}, along with a comparison of the tangential velocity with the radial velocity.  For an isotropic stellar velocity distribution, the tangential velocity should be $\sqrt{2}$ times the radial velocity distribution, since it has two velocity components. Our stars cluster around the isotropic distribution and only 1 star has $v_{\rm tan}>5\times v_{\rm rad}$.

\subsection{Hypervelocity star candidates}

We define the probability of being unbound ($p_{\rm unbound}$) as the fraction of 1000 random realizations of $v_{\rm grf}$, drawn assuming Gaussian uncertainties in the proper motions, distances, and radial velocities, that exceed the local escape velocity at the corresponding position. $7$ of the stars in our sample are potential hypervelocity stars with the probability of being unbound greater than $50\%$. Two stars have a $100\%$ probability. $17$ more stars have a $>50\%$ probability of being unbound assuming a main-sequence nature. However, since the $v\sin i$ of these stars is lower than $50$ \kms\, they could be BHBs instead. This criterion has been used to separate out BHB and MS star candidates by \citet{2025A&A...704A.326C}. A low inclination of the rotational axis could, however, also account for the low projected rotational velocity. Nevertheless, we repeat the procedure described in Section \ref{sec:three} and calculate velocities assuming a typical HB mass of $0.5$ \msun\, \citep[similar to][]{2025A&A...704A.326C}. None of the stars are hypervelocity stars in this case. The results are summarized in Table~\ref{tab:unbound_vgrf}.

\begin{table}[h]
\centering
\setlength{\tabcolsep}{2pt}  
\renewcommand{\arraystretch}{1.1}
\caption{Unbound candidates sorted by descending Galactocentric rest-frame velocity.}
\label{tab:unbound_vgrf}
\begin{tabular}{r c c c c c c}
\hline
\# &
$v_{\rm grf}^{\rm MS}$ &
$v_{\rm grf}^{\rm BHB}$ &
$p_{\rm unbound}^{\rm MS}$ &
$p_{\rm unbound}^{\rm BHB}$ &
$R_{\rm sph}^{\rm MS}$ &
$v\sin i$ \\
&
[km s$^{-1}$] &
[km s$^{-1}$] &
[\%] &
[\%] &
[kpc] &
[km s$^{-1}$] \\
\hline
\multicolumn{7}{c}{$v\sin i \ge 50~\mathrm{km\,s^{-1}}$} \\
\hline
1  & $936 \pm 123$ &- & 100 & - & $100 \pm 11$ &$89^{+19}_{-19}$ \\
5 & $705 \pm 55$   & -   & 100  & -  & $42 \pm 4$   & $55^{+10}_{-11}$ \\
2  & $604 \pm 118$  & - & 99  & -  & $94 \pm 10$  & $62^{+18}_{-19}$ \\
7 & $585 \pm 68$   & -  & 78  & -  & $23 \pm 4$   & $127^{+8}_{-12}$ \\
3 & $450 \pm 83$   & -  & 56  & -  & $44 \pm 5$   & $161^{+13}_{-22}$ \\
4 & $435 \pm 50$   & -   & 63  & -  & $68 \pm 10$  & $57^{+16}_{-18}$ \\
6 & $440 \pm 48$   & -   & 52  & -  & $42 \pm 4$   & $250^{+0}_{-5}$ \\
\hline
\multicolumn{7}{c}{$v\sin i < 50~\mathrm{km\,s^{-1}}$} \\
\hline
11 & $1257 \pm 266$\tablefootmark{a} & $337 \pm 52$\tablefootmark{a}   & 100 & 0   & $36 \pm 8$   & - \\
8  & $857 \pm 152$  & $264 \pm 58$   & 100 & 1   & $114 \pm 14$ & - \\
9 & $848 \pm 83$   & $177 \pm 35$   & 100 & 0   & $66 \pm 8$   & - \\
13 & $806 \pm 89$   & $164 \pm 25$   & 100 & 0   & $71 \pm 7$   & - \\
10 & $760 \pm 86$   & $286 \pm 33$   & 98  & 0   & $36 \pm 5$   & - \\
18 & $717 \pm 80$   & $210 \pm 40$   & 99  & 0   & $25 \pm 2$   & - \\
12 & $699 \pm 112$  & $150 \pm 24$   & 98  & 0   & $50 \pm 7$   &- \\
17 & $551 \pm 61$   & $253 \pm 33$   & 83  & 0   & $27 \pm 3$   & - \\
16 & $531 \pm 59$   & $240 \pm 24$   & 80  & 0   & $41 \pm 5$   &- \\
14 & $508 \pm 54$   & $194 \pm 26$   & 79  & 0   & $35 \pm 4$   & -\\
15 & $466 \pm 40$   & $118 \pm 18$   & 64  & 0   & $39 \pm 3$   & - \\

\hline
\end{tabular}
\tablefoot{Index of the star corresponds to the index in the final table given in the Appendix. For stars with $v\sin i<50$ \kms\, two Galactocentric rest-frame velocities are computed. $R_{\rm sph}$ is the Galactocentric radius of the star at present assuming MS nature. Uncertainties on $v\sin i$ are statistical. \tablefoottext{a}{With parameters derived using the BHB/ELM WD grid. See Section 6.1 for a discussion on the nature of this object and the possibility that it is most likely a pre-ELM WD candidate.}}
\end{table}

\begin{figure*}
   \centering
   \includegraphics[width=0.89\textwidth]{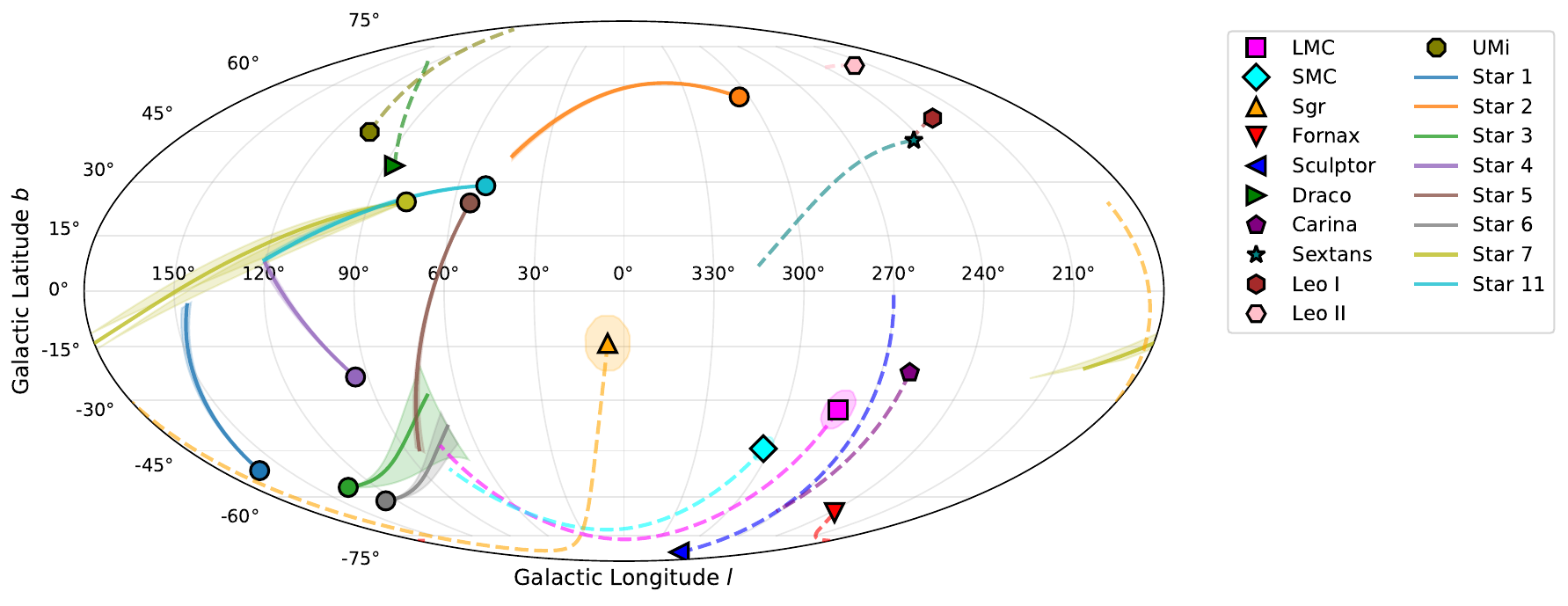}
   \caption{Past trajectories of the hypervelocity star candidates which have $v\sin i>50$ \kms, and LMC in the Mollweide projection in Galactic coordinates. The effect of LMC's gravity has been taken into account. Stellar trajectories are plotted for the $84^{\rm th}$ percentile of age while satellite galaxies are plotted for the maximum of these ages ($400$ Myr). Shaded regions show trajectories which assumed a distance taking the $1\sigma$ distance uncertainties of the stars into account. }
    \label{fig:LMCtraceback}%
    \end{figure*}

Out of these objects only 1 (Gaia DR3 2123475575386817920) crosses the disc within a Galactic radius of $25$ kpc within a flight-time compatible with the $1\sigma$ uncertainty in stellar age. This strongly suggests that the other stars are either high-velocity halo outliers or were ejected either from a satellite galaxy or from a globular cluster. For our sample, we traced back the trajectories for the star along with known satellite galaxies including the LMC and the SMC. The satellite galaxy data was taken from \citet{2022ApJ...940..136P} and \citet{2018A&A...616A..12G}. The LMC and SMC proper motions, RVs, and distances are from \citet{2013ApJ...764..161K}, \citet{2006AJ....131.2514H}, \citet{2002AJ....124.2639V}, and \citet{2019Natur.567..200P}, while the masses are from \citet{2019MNRAS.487.2685E} and \citet{2020MNRAS.497.4162V}. 2D plots (which are often used in similar studies) show 2 overlapping stars as shown in Fig.~\ref{fig:LMCtraceback}, but our 3D traceback, which considered the moving potential of the satellites, did not show them coming from the same point. 

\paragraph{Gaia\,DR3\,2515789291339579648}
The most extreme candidate star in our study has a median S/N of only $13$ and has a single exposure available in the DESI database. No variability is seen in the ZTF light curve, and the SED shows no evidence of a composite system. Its position suggests it is moving toward us. If confirmed as a hypervelocity star (HVS), it could indicate the presence of another dwarf galaxy in the southern hemisphere. This object therefore represents a prime target for spectroscopic follow-up. Our ADS grid classifies this star as metal-poor (\feh $\leq -1$). However, this value falls at the grid boundary for a low S/N spectrum, so the estimate should be treated with caution. We use low metallicity (\feh\, $=-1$) MIST tracks for this star. Since the star may have an even lower metallicity, the distance and velocity estimates are upper bounds.

\paragraph{Gaia\,DR3\,4584485956183915776}

This star also has only 1 exposure, but with a S/N of 37. Although the star has crossed the Galactic plane in the past, this crossing is at $48.5\pm5.1$ kpc, where the density of the disc is almost two orders of magnitude lower than at the solar circle. Star-formation is only happening within $~10-15$ kpc of the GC \citep{2026A&A...708A.252F} and even the spiral arms seem to extend up to $\sim25$ kpc \citep{2015ApJ...801..105X}. Therefore, this star was most likely not ejected from the disc.

\paragraph{Gaia\,DR3\,3677213628601638656}

We downloaded the individual exposures of the star which resulted in five spectra, each with a S/N between 13 and 20. No significant RV variation was detected within $\pm20$ \kms\ implying that the star most likely does not have a close unseen companion. 

\paragraph{Gaia\,DR3\,2123475575386817920}

This star was likely ejected from the disc $\sim41$ Myr ago from a Galactic radius of $\sim25$ kpc.

\paragraph{Gaia\,DR3\,1890511670168860416}
This star was flagged as a DA white dwarf using SDSS photometry by \citet{2011MNRAS.417.1210G}. While the \teff\, reported by them matches ours at $\sim13000$ K, the \logg\, of this star ($3.72\pm0.1$ dex) is much lower than their photometric \logg\,. Our \logg\, shows that it is definitely not a WD.

\paragraph{Gaia\,DR3\,2564538681977454208}
This star is in the BHB catalog of \citet{2021A&A...654A.107C}, based on their reduced proper motion cut. The $v\sin i$ of $161^{+12}_{-22}$ \kms\, however, suggests a MS characteristic. This star has 2 exposures with S/N of 13 and 14. No RV variability is detected between these two exposures. This star is also moving towards the disc.
\paragraph{Gaia\,DR3\,2537649197407027584}

This star has the highest $v\sin i$ in the sample ($v\sin i>250$ \kms), and is a MS star. It was studied by \citet{1992MNRAS.257..225A}, but they could not get a valid distance estimate. They found a heliocentric RV of $-434$ \kms, lower than our estimate of $\sim-370$ \kms. Two exposures of this star in DESI, however, were non-variable in RV. 

\paragraph{Gaia\,DR3\,4595082297634366592}
This star (\#11) may be the fastest star in our sample if we assume an MS nature. This star was initially classified as a ``narrow-line hydrogen star" by \citet{2015MNRAS.448.2260G} and as an sdB by \citet{2016MNRAS.455.3413K}. It appears in the known hot subdwarf catalogue of \citet{geiersdb}. Our spectral results do not show evidence for being an sdB and are more in line with the classification of \citet{2018MNRAS.475.2480P}, who prefer a MS nature slightly over a pre-ELM WD). 

We find that the RV of this star ($-233\pm 11$ \kms) is similar to that derived using the SDSS spectra ($-242\pm4$ \kms\,). Our MS B-type model-fit has \logg\, $\sim3.8$ and $v\sin i\sim80$ \kms\,, while our BHB/ELM-WD model-fit has \logg\, $\sim 4.0$ and $v\sin i<50$ \kms\,. If this star is an MS B-type star then it is significantly farther away at $\sim 50$ kpc. However, this star does not cross the Galactic plane at any point. Consequently, no known mechanism can be invoked to accelerate an MS star to such high velocities outside of the disk. This velocity is significantly higher than the velocities of any extra-galactic HVS candidate. We consider this star unlikely to be an MS star and more likely to be a pre-ELM WD with $V_{\rm}\sim 200$ \kms\,. A high-resolution spectrum is required to measure metal abundances and $v\sin i$ to differentiate between pre-ELM WD and MS nature.

\subsection{Runaway stars}

\begin{figure}
\centering
    \includegraphics[width=0.47\textwidth]{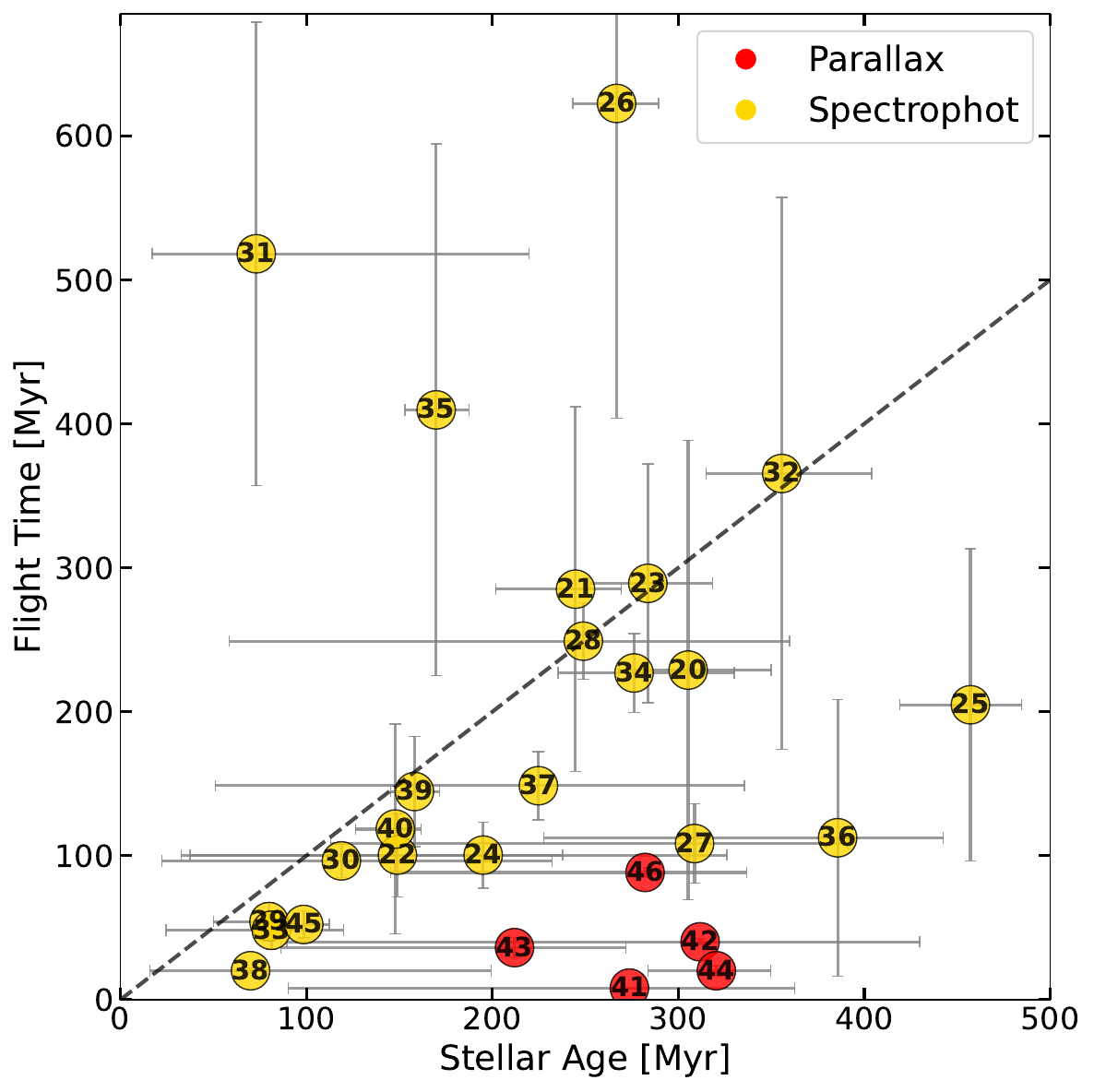}
    \caption{The time of last crossing of the stellar disc as a function of the stellar age from MS evolutionary tracks for our sample. The dashed grey line and the corresponding shaded regions show the regime wherein the flight time and stellar age overlap within uncertainties. Red stars are stars with distances derived from \gaia\, parallax.}
    \label{fig:flightimes}
\end{figure}

Our sample contains $27$ stars that are most likely bound runaway stars ejected from the disc. Their metallicities and total velocities are in line with disc ejections, but higher-resolution follow-up spectra will allow these stars to be confirmed as runaway stars. Of these, 26 stars have non-zero flight-times consistent with their stellar ages at the $2\sigma$ level, while 23 of these agree at the $1\sigma$ level. Gaia\,DR3\,1224962262373806976 is the only one where the flight time of $\sim530$ Myr is significantly larger than from the age of $\sim73$ Myr. The flight times of the runaway star candidates are shown as a function of the stellar age in Fig.\,\ref{fig:flightimes}.  Stars close to the $1:1$ line are prime candidates for a dynamical ejections whereas stars with stellar ages significantly higher than the flight times were most likely produced in the BSS. Due to the high uncertainties in the distances of these stars it is not possible to find clusters of ejections for these stars.
Like with the HVS candidates of Section 6.1, the projected rotational velocity can allow us to identify bona-fide runaway candidates. 13 out of 27 stars have a projected rotation velocity $v\sin i<50$ \kms\, within $1\sigma$ uncertainty. 7 of these are in the over-density between the $2$ \msun\, and $3$ \msun\, evolutionary tracks in Fig.\ref{fig:kielwithtracks}. The final distribution of candidates is given in Table~~\ref{tab:summary}.

Five runaway stars have reliable parallaxes, and the parallax based mass of these stars confirms the mass through evolutionary tracks (see Appendix A). In Fig.~\ref{fig:ejectionvel} the ejection velocity of the runaway candidates is plotted as a function of the radius of Galactic plane-crossing. A scale height of $1$ kpc is assumed to propagate uncertainties for the time-of-flight.  Gaia\,DR3\,1488588768046793600, Gaia DR3 3791270299077640832, Gaia\,DR3\,4551328950397878656, and Gaia\,DR3\,5758776812872962560 are consistent with a Galactic centre or bulge origin, while Gaia\,DR3\,4573060866436078208 is more likely associated only with the bulge. 

The fiducial binary population synthesis (BPS) model of \citet{2020MNRAS.497.5344E} demonstrate that producing runaway stars 
with velocities exceeding $\sim200$ \kms\, is highly improbable, requiring heavy optimization of initial parameters to surpass this threshold. This fiducial limit is independent of the choice of population synthesis code \citep{2025OJAp....8E..85W}, since the ejection velocity of a runaway star is primarily determined by its pre-supernova orbital velocity. Our sample is more consistent with an origin in the dynamical ejection scenario (DES) or the two-step ejection mechanism. A subset of our runaway stars exhibits high ejection velocities ($>450$ \kms\,), placing them beyond the usual range expected from existing models of runaway star formation. However, none of our runaways are as extreme as the fastest runaways of \citet{irrgang21}, and can still be explained by dynamical ejections in extreme cases. Ejection velocities of $300-540$ \kms\, were produced by runaways undergoing multiple dynamical processes in the $n$-body simulations of \citet{2012ApJ...751..133P} and \citet{schoettler2020}.

The upper panel of Fig.~\ref{fig:ejection2} presents ejection velocities as a function of Galactic plane-crossing radius, with fast disc runaways from \citet{2021A&A...646L...4I} and \citet{Raddi2021} included for comparison (these samples followed a similar procedure as outlined in this paper). The disc density profile from Model~I of \citet{andreas}, used to compute the stellar trajectories, is shown as the purple curve. 
With the exception of four stars from \citet{2021A&A...646L...4I}, stars ejected at velocities exceeding $450$\,\kms\ are preferentially associated with the inner disc, suggesting a correlation with disc density. A denser disc environment with ongoing star formation may promote the formation of denser open clusters \citep[see for example][]{2019MNRAS.482.2530R}, potentially allowing ejection velocities beyond the limits predicted by classical cluster ejection simulations. As shown in the lower panel of Fig.~\ref{fig:ejection2} hyper-runaways are formed at almost all masses. \citet{2012ApJ...751..133P} predict a population of hyper-runaways between $2-4$ \msun\, which, however, is lower in fraction than stars with masses $>4$ \msun. An exact dependence of ejection velocity on mass would require a bigger sample of runaways and should be considered in future studies. We also note here that one star in the sample is outside of a plane-crossing radius of $50$ kpc within uncertainties. Star-formation and spiral arms are not seen at those limits, meaning that there is no standard way to eject a runaway star there (see for example, the discussion on Galactic disc size-limits in \citet{irrgang21}). This star is therefore not a disc runaway.

We find that there is also a region of avoidance for fast intermediate-mass runaways, similar to what was found by \citet{sana2022} ($v_{\rm los}>60$ \kms), and more recently by \citet{2026A&A...705A.215C} ($v^{2D}_{\rm pec}>85$ \kms) for O-type stars. Almost no fast runaways have $v\sin i>200$ \kms\,. We find only 1 runaway star (Gaia\,DR3\,5758776812872962560) which shows rotation consistent with  $v \sin i >200$ \kms\, in this sample. Since our sample is located exclusively in the faster runaway space we re-confirm the lack of fast rotators in this region of avoidance.
\begin{table}
\centering
\caption{Summary of the sample classification.}
\label{tab:summary}
\begin{tabular}{lc}
\hline\hline
Category & Number \\
\hline
Total sample & 54 \\
Below ZAMS & 8 \\
HVS candidates ($v\sin i > 50$\,km\,s$^{-1}$) & 8 \\
HVS candidates ($v\sin i < 50$\,km\,s$^{-1}$) & 11 \\
Runaway candidates ($v\sin i > 50$\,km\,s$^{-1}$) & 14 \\
Runaway candidates ($v\sin i < 50$\,km\,s$^{-1}$) & 13  \\
\hline
\end{tabular}
\end{table}

\begin{figure}
    \centering
    \includegraphics[width=0.5\textwidth]{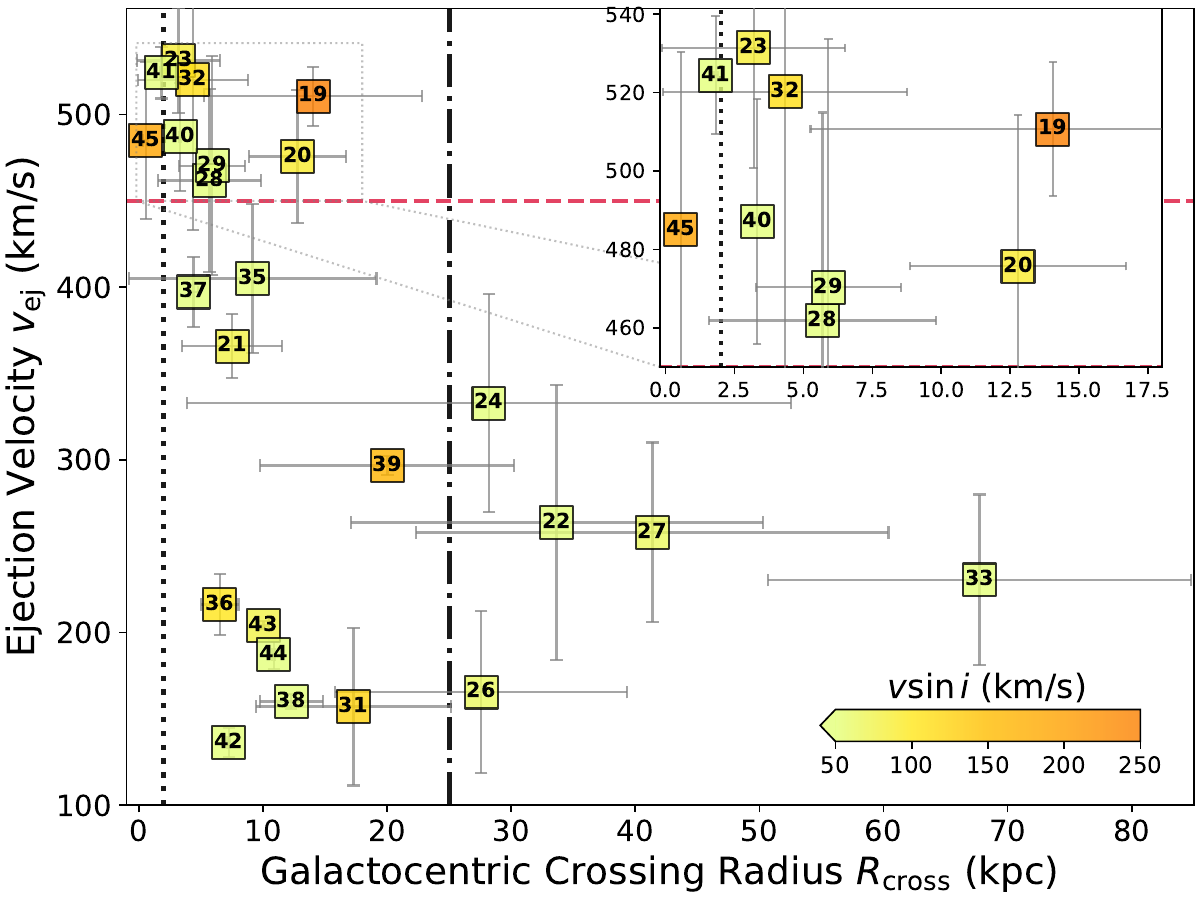}
    \caption{Ejection velocities of the runaway candidates as a function of the radius of Galactic plane crossing. The black dotted line (dash-dotted) marks the inner 2 (25) kpc of the disc and the red dashed line marks an ejection velocity limit of $450$ \kms. 9 stars are above the ejection velocity limit. The colourbar shows the $v\sin i$ of the stars.}
    \label{fig:ejectionvel}
\end{figure}

\begin{figure}
    \centering
    \includegraphics[width=0.49\textwidth]{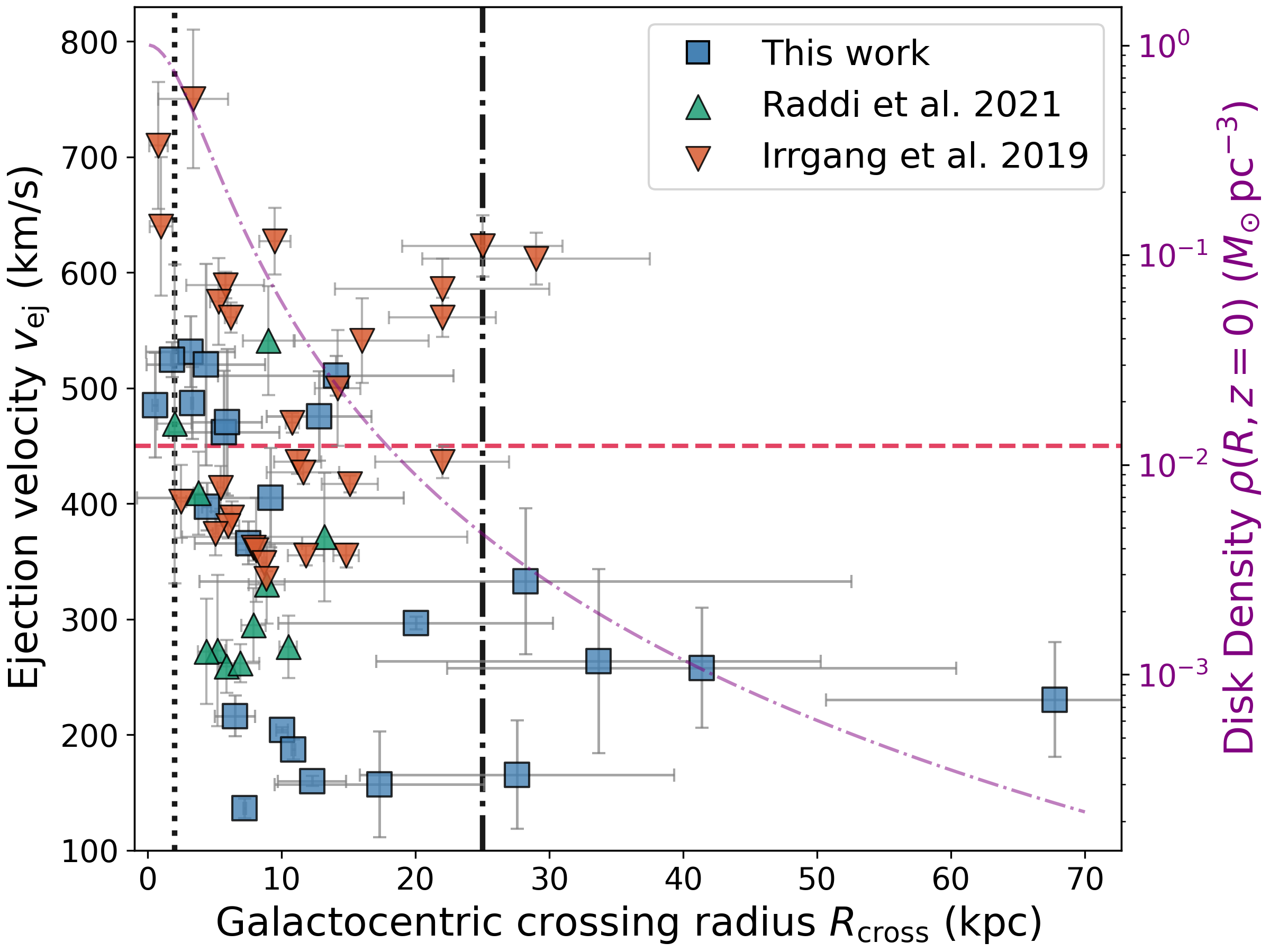}
    \includegraphics[width=0.49\textwidth]{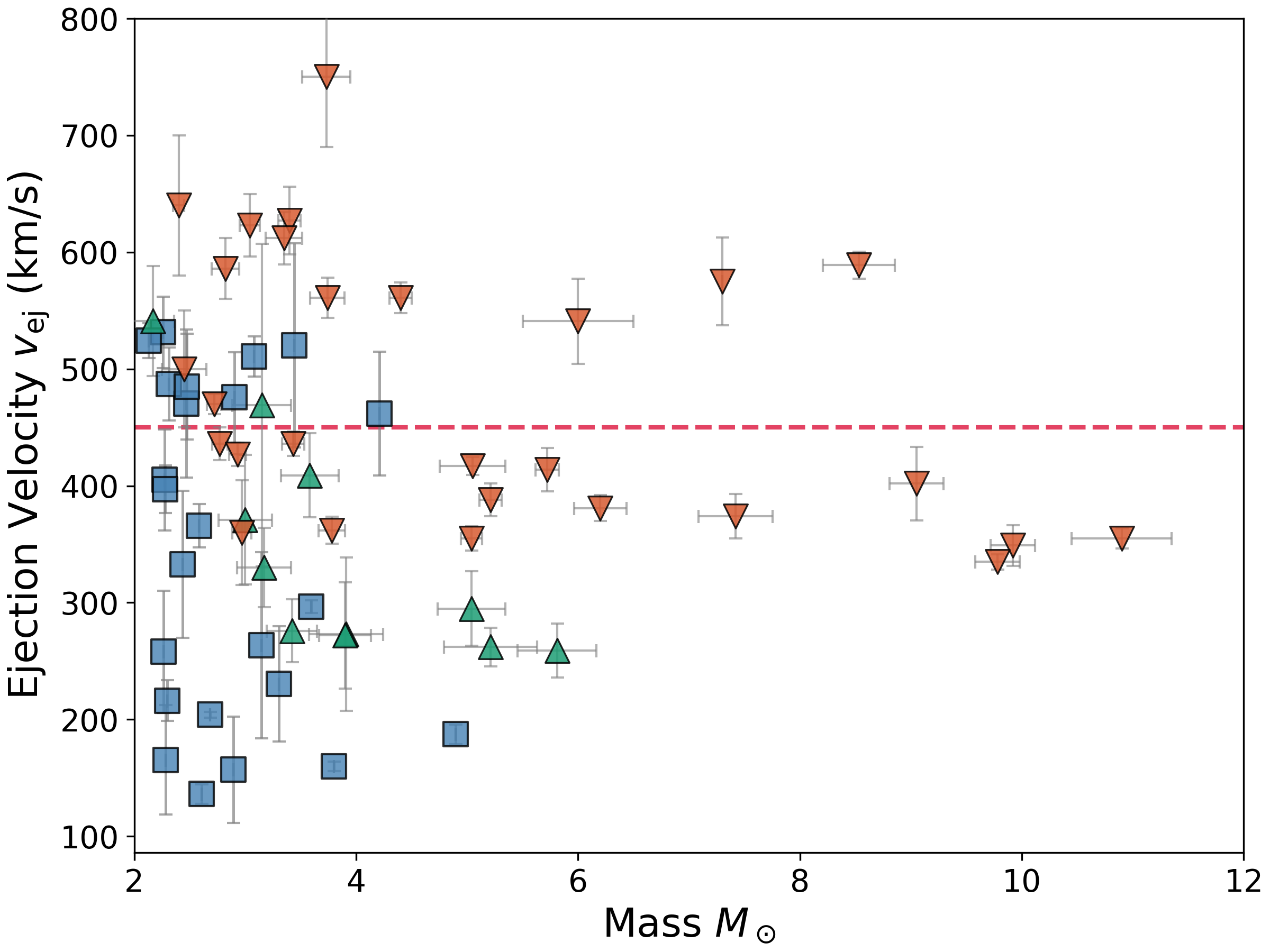}
    \caption{Top: Same as Fig.~\ref{fig:ejectionvel} but including MS candidates from \citet{2021A&A...646L...4I} (orange upside-down triangles) and MS candidates from \citet{Raddi2021} (green triangles). Purple line shows the density of the disc in the potential of \citet{andreas} in log-scale, with an increased density most likely correlated to higher ejection velocities through the DES. Bottom: Ejection velocities as a function of mass. The three fast candidates above $6$ \msun\, from \citet{2021A&A...646L...4I} were all formed in the inner disc.}
    \label{fig:ejection2}
\end{figure}

\section{\gstarDESIshort: An extreme runaway star showing light curve variability}

The star with the highest measured RV in our sample of $513\pm20$ \kms\, is Gaia\,DR3\,3573377850714810496. We acquired another spectrum of the star at the Nordic Optical Telescope (NOT) using the ALFOSC instrument, to check for RV variability. No RV variability was detected. The DESI spectrum with visible hydrogen and helium lines is shown in Fig.~\ref{fig:spectra}, and the atmospheric parameters are given in Table~\ref{tab:params1}. The star is a metal-rich B-type MS star with a helium abundance of $-0.8\pm0.1$, although we do not see any evidence of fast rotation, which implies that the star could be slowly rotating.
\begin{figure}
   \centering
   \includegraphics[width=0.5\textwidth]{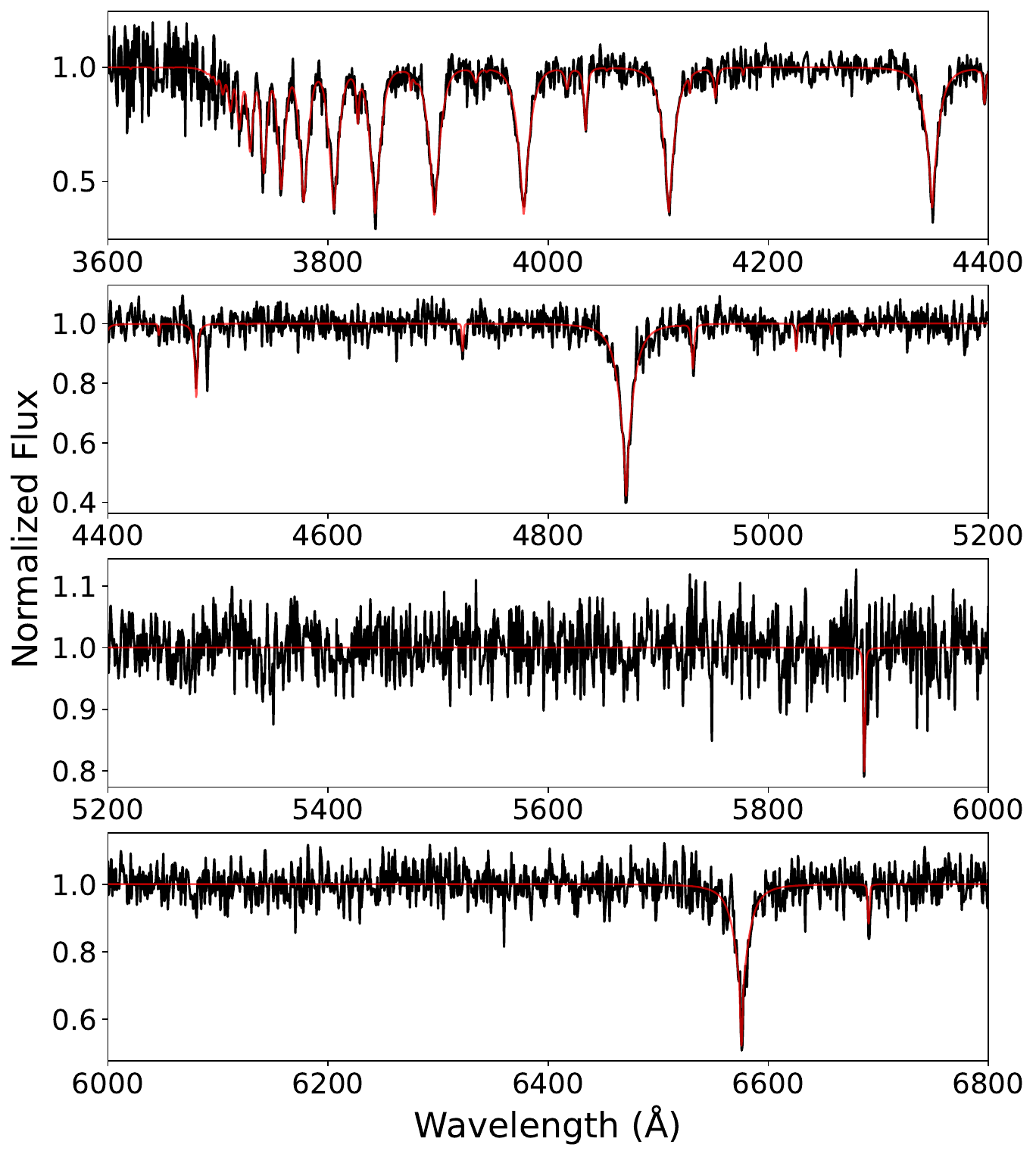}
   \caption{DESI spectrum of \gstarDESIshort\,(black) and model fit (red). Hydrogen Balmer and helium lines are clearly visible.}
    \label{fig:spectra}%
    \end{figure}
    
\begin{table}
\renewcommand{\arraystretch}{1.1}
\centering
\caption{Best fit spectro-photometric and MIST parameters. Statistical uncertainties are provided for all the quantities.}
\begin{tabular}{c c c}
\hline
\hline
Parameter& \gstarDESIshort\\
\hline
$T_{\mathrm{eff}}$ [K]& $14392^{+300}_{-314}$  \\
log $g$ & $4.03 \pm 0.10$  \\
 \text{[Fe/H]} & $0.3 \pm 0.2$ \\
$v\sin i$ [km\,s$^{-1}$] & $<50$ \kms \\
$v_{r}$ [km\,s$^{-1}$] & $513 \pm 11$ \\
\hline
Angular diameter $\log \Theta$ & $-11.273 \pm 0.006$ \\
Colour excess $E(44-55)$& $0.006^{+0.008}_{-0.006}$  \\
Evolutionary mass [\msun]& $4.06\pm 0.10$  \\
Radius ($R=\sqrt{GM/g}$) [$R_\odot$]  & $3.05 \pm 0.09$\\
Distance d$_{\rm spectro}$ [kpc] & $24^{+10}_{-5}$ \\
Luminosity [$L_\odot$]& $325\pm 34$ \\
\hline
\end{tabular}
\label{tab:params1}
\end{table}

The SED fit is shown in Fig.~\ref{fig:SED}, along with an evolutionary track fit, shown in the upper panel of Fig.~\ref{fig:MIST}. The star is close to $90$ Myr old and is still in the main sequence phase with a mass of $4.06\pm0.10$ \msun. The posterior of the distribution is shown in the lower panel of Fig.~\ref{fig:MIST}. 

\subsection{Kinematics}
\begin{figure}
\centering
    \includegraphics[width=0.48\textwidth]{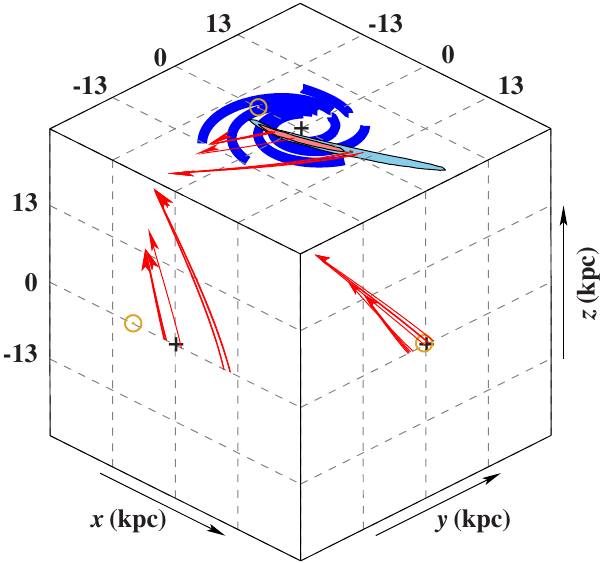}
    \caption{9 example trajectories for the star \gstarDESIshort. Sun is the yellow circle and the Galactic centre is the black cross. $2\sigma$ contours are shown for the positions of disc crossing showing that the star most likely comes from the inner disc. Blue arcs are spiral arms.}
    \label{fig:kinematics}
\end{figure}

We plot a sub-sample of 9 representative trajectories of this star in Fig.~\ref{fig:kinematics}. The spiral arms are plotted as blue arcs using the polynomial functions of \citet{2014A&A...569A.125H}. The star was ejected from the disc $45^{+23}_{-10}$ Myr ago with an ejection velocity of $442^{+79}_{-9}$ \kms. The closest radial distance was $3.6$ kpc. The difference between the evolutionary and kinematic age lends credence to this star being ejected through a binary system in which the primary underwent a supernova. The difference between the stellar age and the kinematic age allows us to put an upper bound on the lifetime of the pre-supernova companion, which was around 65 Myr. For roughly $16\%$ of the trajectories the star is unbound, which is mainly due to the uncertainty on the distance. Therefore, the stars is most likely bound. Despite the distance uncertainty, a Galactic centre ejection can be discarded, and an inner disc ejection is favoured, which matches the slightly higher metallicity of the star. 

\begin{figure}
    \centering
    \includegraphics[width=0.48\textwidth]{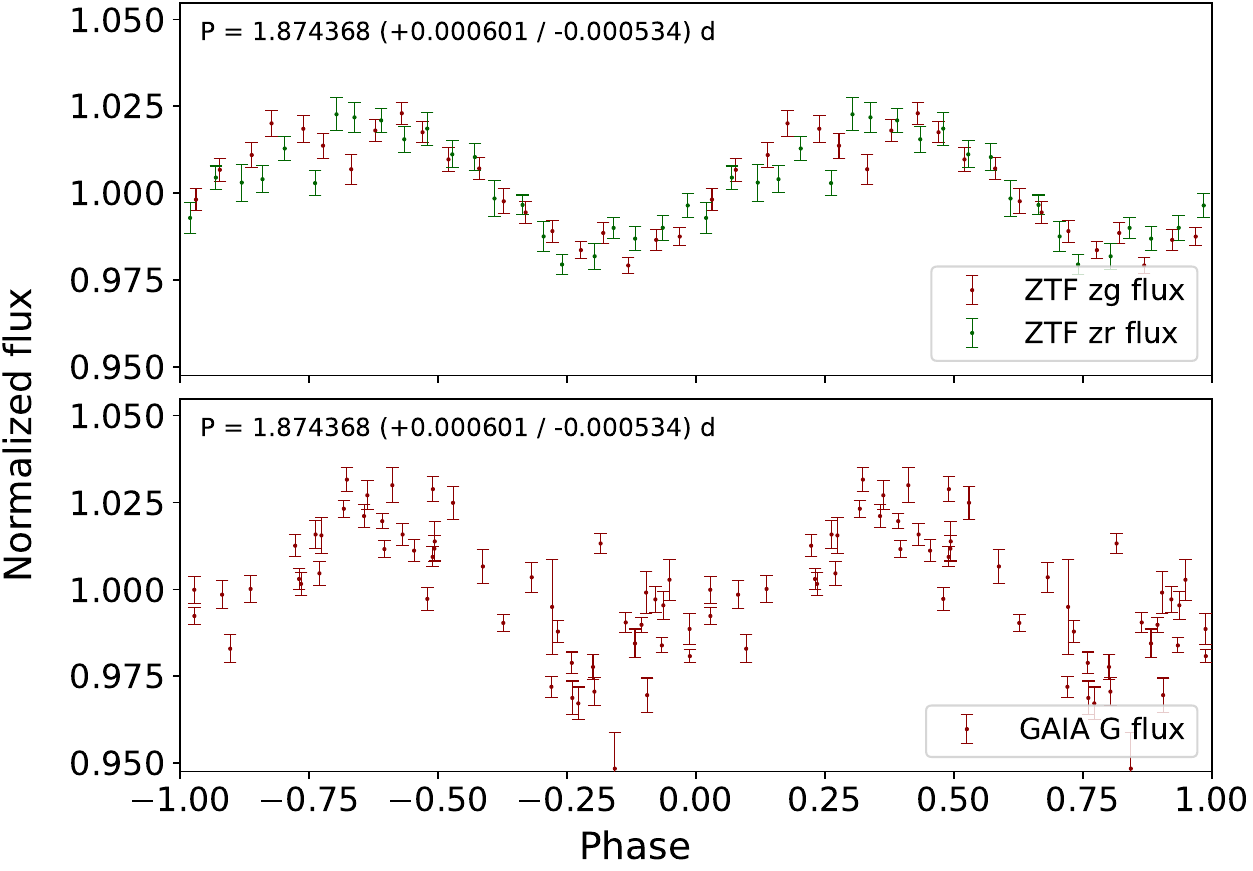}
    \caption{ ZTF (Upper panel) and Gaia g-band (lower panel) light curves for \gstarDESIshort\ phase-folded to a period of $1.8743$ days.}
    \label{fig:LC}
\end{figure}

\subsection{Light curve}
We retrieved the \gaia\, light-curve of the star which is quite noisy. The \gaia\ variability catalogue flags it as an eclipsing binary with a period of $\sim3.7$ days \citep{Gaia3}. This star also has a light-curve in ZTF. We used both light-curves and create individual generalized Lomb-Scargle Periodograms \citep{2009A&A...496..577Z}. These periodograms are then multiplied together to find a maximal period half of the \gaia\, one ($1.874368\pm0.000601$ days). The phase-folded light curves are shown in Fig.~\ref{fig:LC}\footnote{Light curves were retrieved and phase-folded using the Python script \href{https://github.com/Fabmat1/lightcurvequery}{https://github.com/Fabmat1/lightcurvequery}.}. This period, combined with a lack of any eclipse, and the non-variability of the RV, leads us to believe that it is a slowly pulsating B star (SPB), very similar to the star reported by \citet{Heber2025}. Another hyper-runaway star PG\,1610+062, studied by \citet{irrgang19} was found to be an SPB, had a similar mass, and low rotational velocity. However, rotational modulation can not be ruled out as of now since a rotation velocity lower than $50$ \kms\, could also have a similar modulation period. We consider whether the observed variability could arise from ellipsoidal modulation, wherein a hot main-sequence primary is expected to be tidally synchronized. Assuming synchronous 
rotation at the photometric period, which is likely at this period \citep{2024A&A...688A.141L}, and adopting our derived radius of $R = 3.05\,R_\odot$, we expect an equatorial rotation velocity of 
$v_{\rm rot} \sim 82$ \kms, substantially higher than our measured $v\sin i$ 
limit of $50$ \kms. This, along with a lack of RV variability in the two spectra, disfavours ellipsoidal modulation as the origin of the photometric variability.

\section{Conclusions}
In this paper we have detailed the spectroscopic, photometric, evolutionary, and kinematic analysis of a selection of candidate intermediate-mass young stars in the Galactic halo selected through a cross-match of \gaia\, and DESI-DR1. The spectra of the stars were fit using an automated spectral-fitting pipeline and candidates were selected with an absolute radial velocity cut-off of $150$ \kms\,. We used the \gaia\, parallax, where available, to de-select BHB stars which are often found in the halo and lie in the same phase space in the Kiel diagram. Combining spectral analysis with photometry and evolutionary tracks, distances to the stars were determined for the case when the \gaia\, parallax was not reliable. The distances to these stars were found to lie between $10-100$ kpc. 

We propagated the distances together with the astrometry and the radial velocity to derive an estimate for the probability of the star to be unbound to the Galaxy. We found $18$ stars to be unbound with a probability higher than $50\%$ if a main-sequence nature was assumed. For $11$ of these stars we found no evidence of rotation greater than $50$ \kms\,. Since such stars could be BHBs, and disentangling them from MS HVS candidates is non-trivial \citep{2008ASPC..392..167H}, we derived Galactocentric velocities assuming a BHB nature as well. None of the $11$ stars were found to be unbound then. With the exception of one object, the unbound star candidates do not trace back to the Galactic disc and are most likely of extragalactic origin although kinematic traceback with known dwarf galaxies did not lead to any candidate ejection sites. In the absence of any known acceleration mechanism, these stars could be fast halo outliers. Follow-up studies and better spectra should help constrain this.

We found $27$ runaway candidates, 14 of which showed evidence for projected rotational velocities greater than $50$ \kms. $8$ of these stars have ejection velocities $>450$ \kms. The star with the highest radial velocity, \gstarDESIshort, was also found to be bound to the Galaxy and was most likely ejected from the inner disc $\sim45$~Myr ago. While the difference in age and time of flight suggest the binary supernova scenario as origin, the ejection velocity is far above what theoretical models predict. Follow-up observation revealed no significant RV variability. The star has \gaia\, and ZTF light curves that show modulations with a period of $\sim1.8$~days. While the star could be a slowly pulsating B-type star, rotational modulation (with a projected rotational velocity less than $50$ \kms\,) cannot be ruled out. Our analysis of candidate HVS and runaway stars combining \gaia\ DR3 and DESI DR1 suggests that the upcoming \gaia\ DR4 and DESI DR2 will substantially increase the number of such stars. The full blue sample will also be substantially increased by these data releases. The high-latitude survey area of DESI is exceptionally well-suited for finding runaway stars, so that spectroscopic distances can be combined with precise \gaia\ proper motions to map a larger volume of runaway and high-velocity stars. This will allow us to study the properties of the high-velocity MS population in an unprecedented, statistically significant manner.

\begin{acknowledgements}
A.B. was supported by the Deutsche Forschungsgemeinschaft (DFG) through grant GE2506/18-1, H.D. through grant GE2506/9-2. M.D. was supported by the Deutsches Zentrum für Luft- und Raumfahrt (DLR) through grant 50-OR-2510.  Fitting procedures were done on the clusters of University of Potsdam, and Dr. Karl-Remeis Sternwarte, Bamberg. This work has made use of data from the European Space Agency (ESA) mission {\it Gaia} (\url{https://www.cosmos.esa.int/gaia}), processed by the {\it Gaia}
Data Processing and Analysis Consortium (DPAC,
\url{https://www.cosmos.esa.int/web/gaia/dpac/consortium}). Funding for the DPAC
has been provided by national institutions, in particular the institutions
participating in the {\it Gaia} Multilateral Agreement. This research has made use of the VizieR catalogue access tool, CDS,
Strasbourg, France \citep{vizier2} and the SIMBAD database,
operated at CDS, Strasbourg, France \citep{simbad}. The data presented here were obtained [in part] with ALFOSC, which is provided by the Instituto de Astrofisica de Andalucia (IAA) under a joint agreement with the University of Copenhagen and NOT. This work made use of the following software packages: \texttt{matplotlib} \citep{Hunter:2007}, \texttt{numpy} \citep{numpy}, \texttt{python} \citep{python}, \texttt{scipy} \citep{2020SciPy-NMeth}, and \texttt{galpy} \citep{2015ApJS..216...29B}.

Software citation information aggregated using \texttt{\href{https://www.tomwagg.com/software-citation-station/}{The Software Citation Station}} \citep{software-citation-station-paper,software-citation-station-zenodo}.
\end{acknowledgements}

\bibliographystyle{aa}
\bibliography{references}
\clearpage

\appendix
\nolinenumbers
\section{Photometry and comparison of distances}

\begin{figure}[h]
   \centering
    \includegraphics[width=0.5\textwidth]{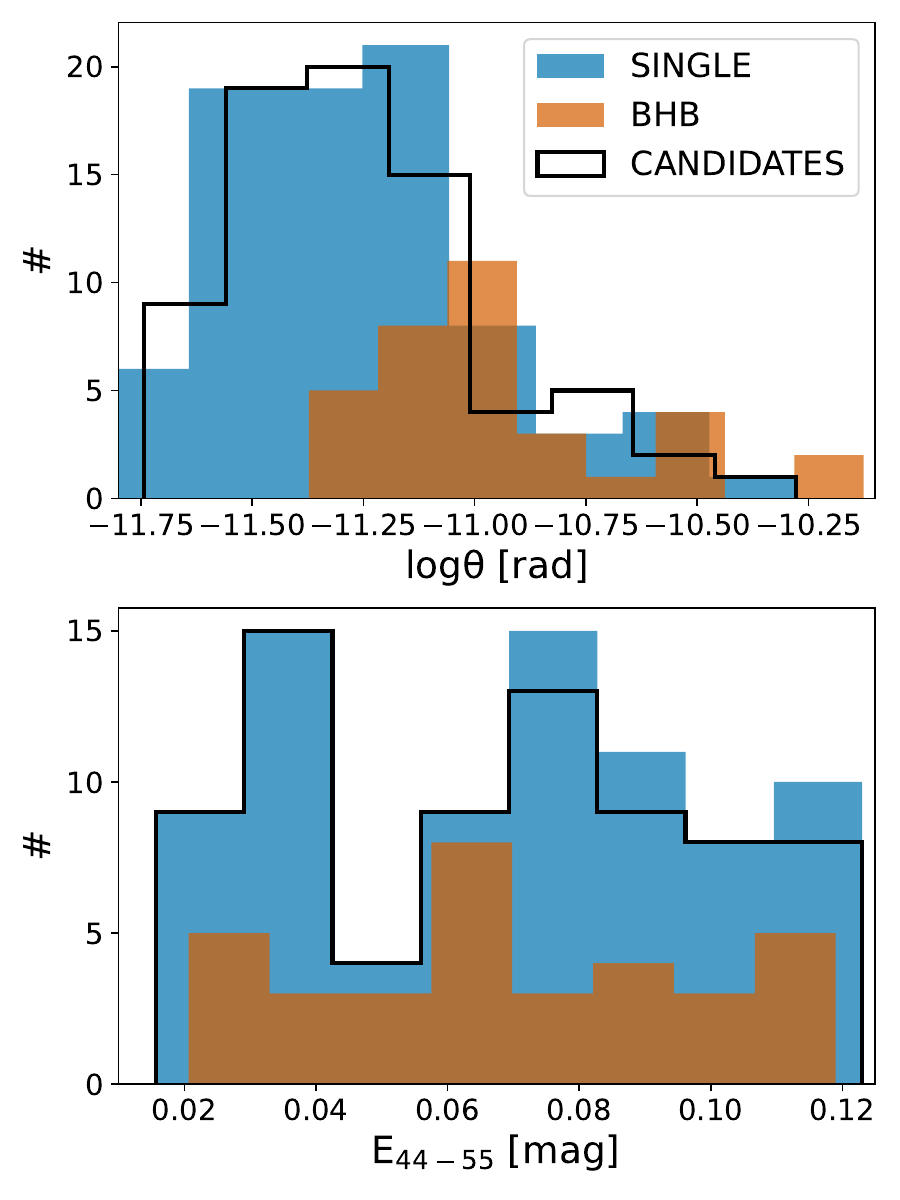}
   \caption{$\log\Theta$ and reddening E(44-55) distribution of the sample. 
   }
              \label{fig:redandtheta}%
    \end{figure}

The angular diameter and redenning (E(44-55)) distributions are plotted in Fig.~\ref{fig:redandtheta}.

\begin{figure}
   \centering
    \includegraphics[width=0.5\textwidth]{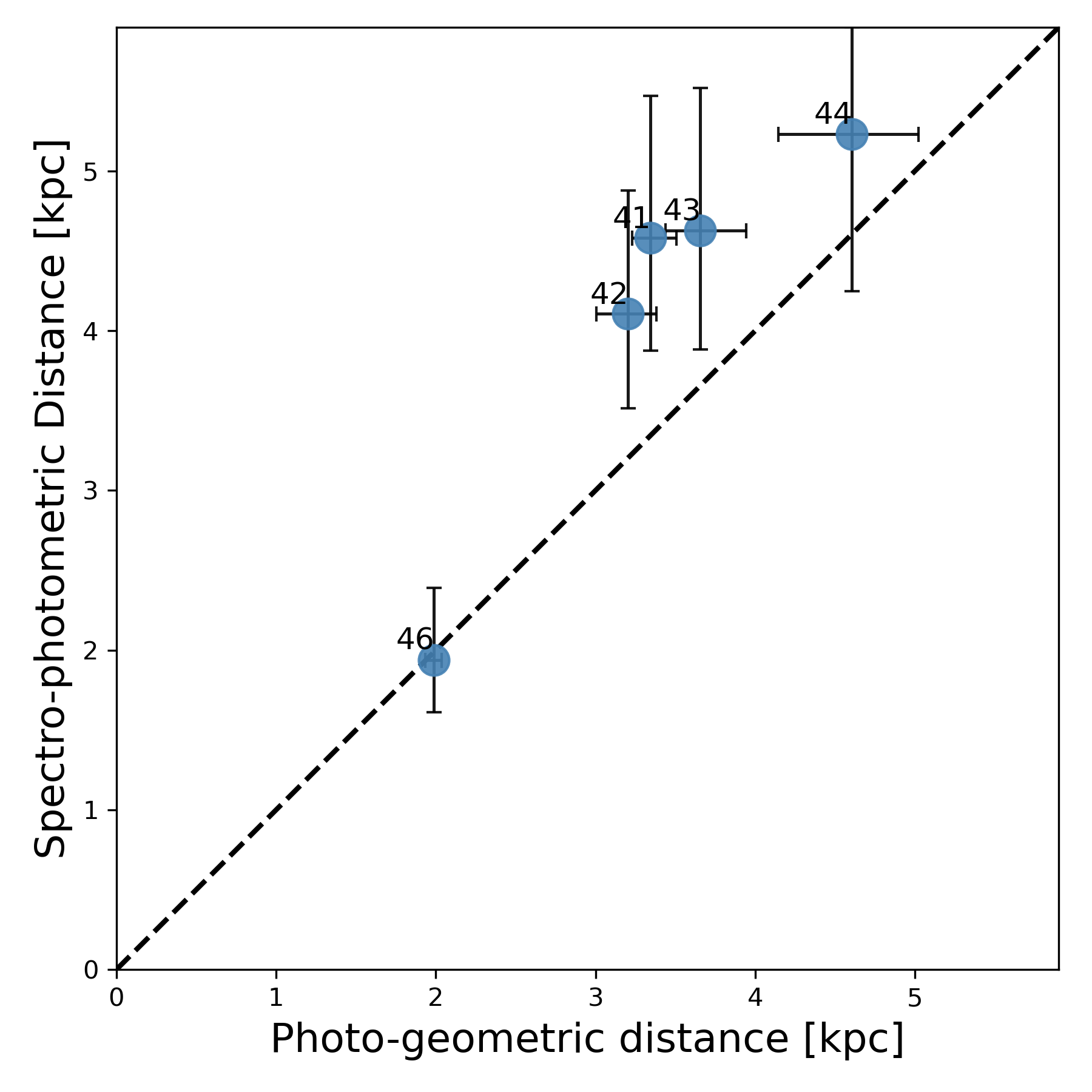}
   \caption{Spectro-photometric distances derived in this work plotted against the photo-geometric distances of \citet{2021AJ....161..147B}.
   }
              \label{fig:distanceandparllax}%
    \end{figure}

For the 5 stars with reliable \gaia\ parallaxes (non-zero parallax and less than 20\% uncertainty) we compare the photo-geometric distances from \citet{2021AJ....161..147B} with the spectro-photometrically derived distances in Fig.~\ref{fig:distanceandparllax}. The uncertainties on the latter are generally more due to the \logg\, uncertainty, with spectro-photometric distances being slightly higher than the photo-geometric one for $3$ stars. 


\section{\gstarDESIshort\, fits}
\begin{figure}
   \centering
   \includegraphics[width=0.5\textwidth]{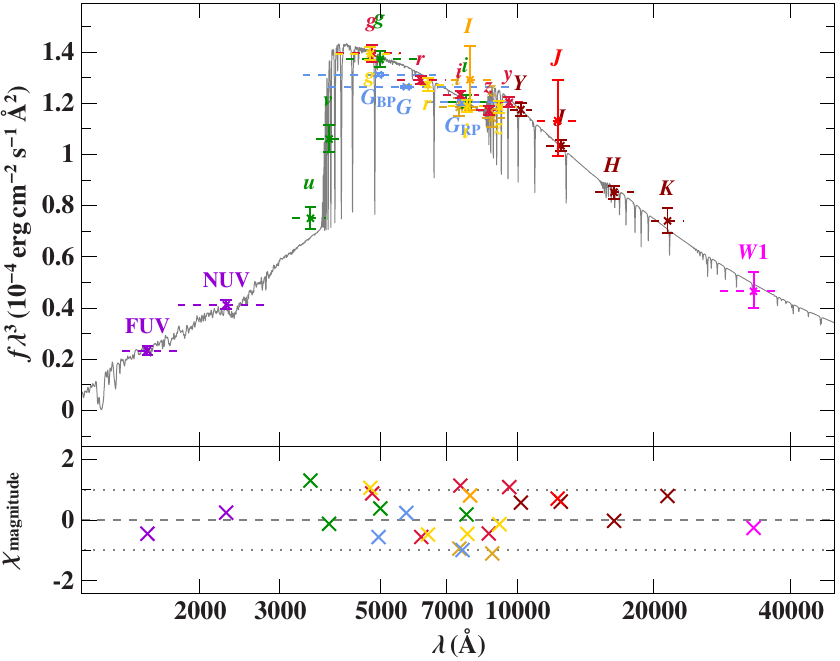}
   \caption{Spectral Energy distribution of \gstarDESIshort. The measurements are taken from \citet{Gaia3}, Wise \citep{wise}, Skymapper \citep{skymapper}, Galex \citep{2005ApJ...619L...1M}, Panstarss \citep{panstarss}, 2MASS \citep{2mass}, DELVE-DR2 \citep{2022ApJS..261...38D}, VST ATLAS \citep{2015MNRAS.451.4238S}, and the VISTA Hemisphere survey \citep{vhs}.}
    \label{fig:SED}
    \end{figure}

\begin{figure}
   \centering
   \includegraphics[width=0.5\textwidth]{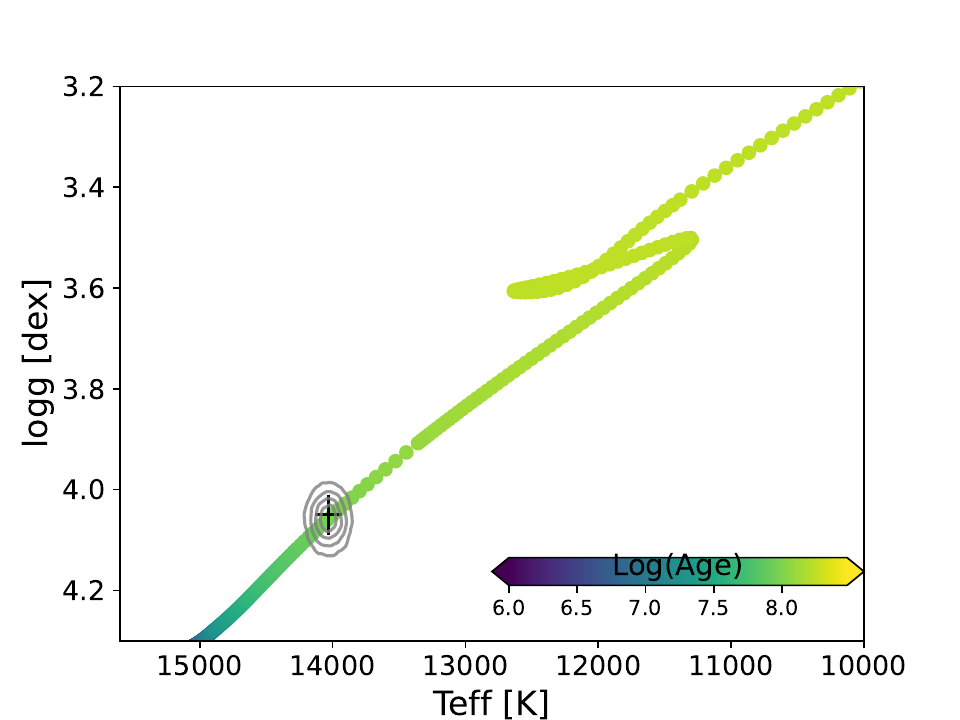}
    \includegraphics[width=0.5\textwidth]{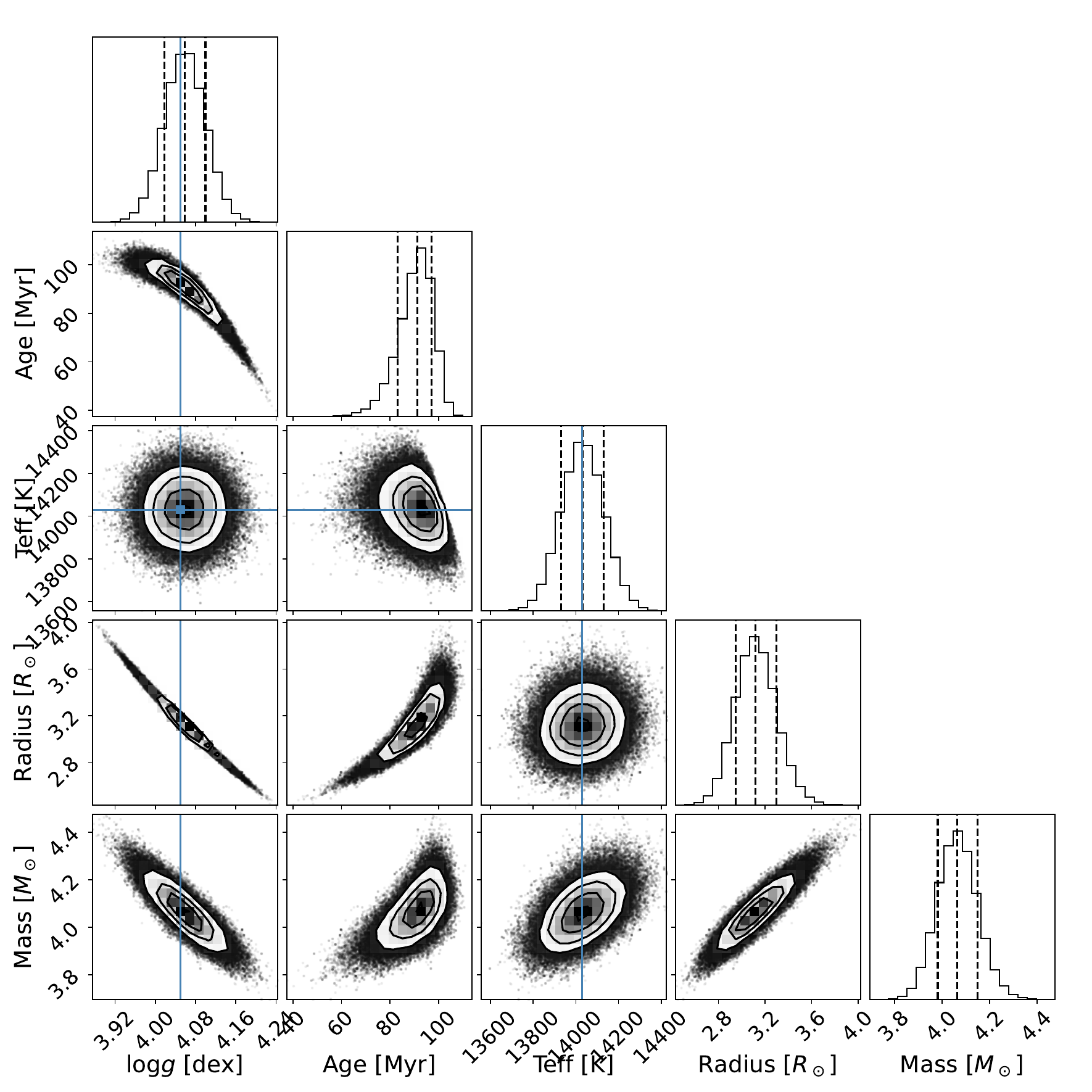}
   \caption{Upper panel: Best-fit MIST tracks for the atmospheric parameters in a Kiel diagram. The contours are the values which are sampled by the Monte Carlo procedure. Lower panel: The corresponding corner plots showing the posterior of the evolutionary parameters.}
              \label{fig:MIST}%
    \end{figure}

The photometric fit for \gstarDESIshort\, is shown in Fig.~\ref{fig:SED}. The evolutionary track and the corresponding corner plots are shown in Fig.~\ref{fig:MIST}.

\section{Spectra}
\label{App:spec}

\begin{figure}[h]
   \centering
   \includegraphics[width=0.5\textwidth,keepaspectratio]{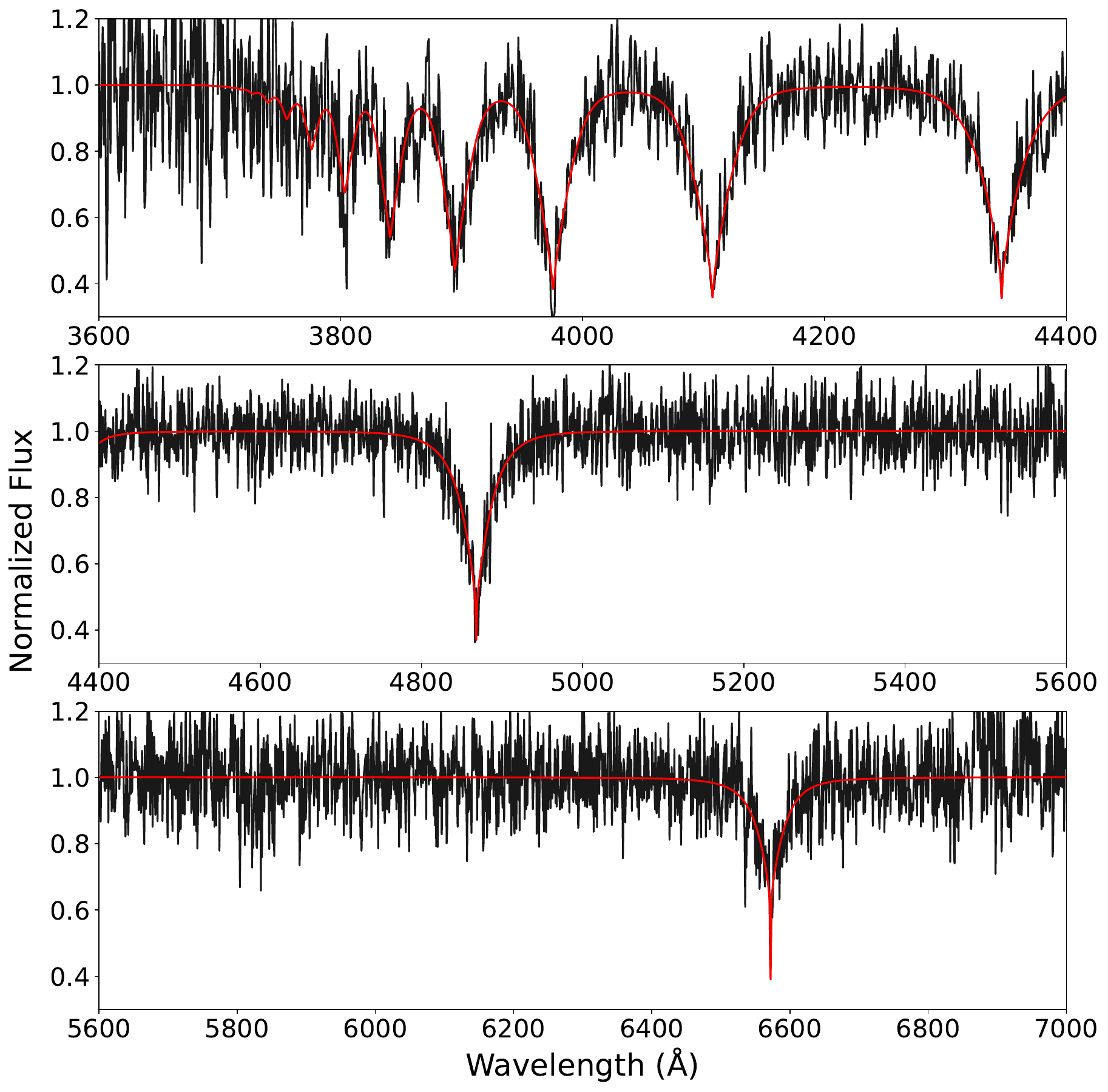}
   \caption{Best model fit (red) to the DESI spectra of the star SBSS 1434+503. We only show the bluer region where the Balmer lines are visible for illustration.}
              \label{fig:examplefit}%
    \end{figure}

The DESI spectrum and best-fit of SBSS 1434+503 is shown in Fig.~\ref{fig:examplefit}. This is an example of an ELM WD star with \teff$=16130\pm 325$ K and \logg$=6.61\pm0.1$. Our best-fit values agree with those of \citet{2026ApJ..1000..216K}, who found $T_{\rm eff}=16919 \pm 407$ and $\log g=6.435\pm 0.115$. The RV of the star in this epoch is $429\pm12$ \kms.
    
The normalized spectra of the HVS candidates with $v_{\rm grf}>600 $ \kms\, is shown in Fig.\ref{fig:spectra}.
 \begin{figure*}
     \centering
     \includegraphics[width=1.0\textwidth]{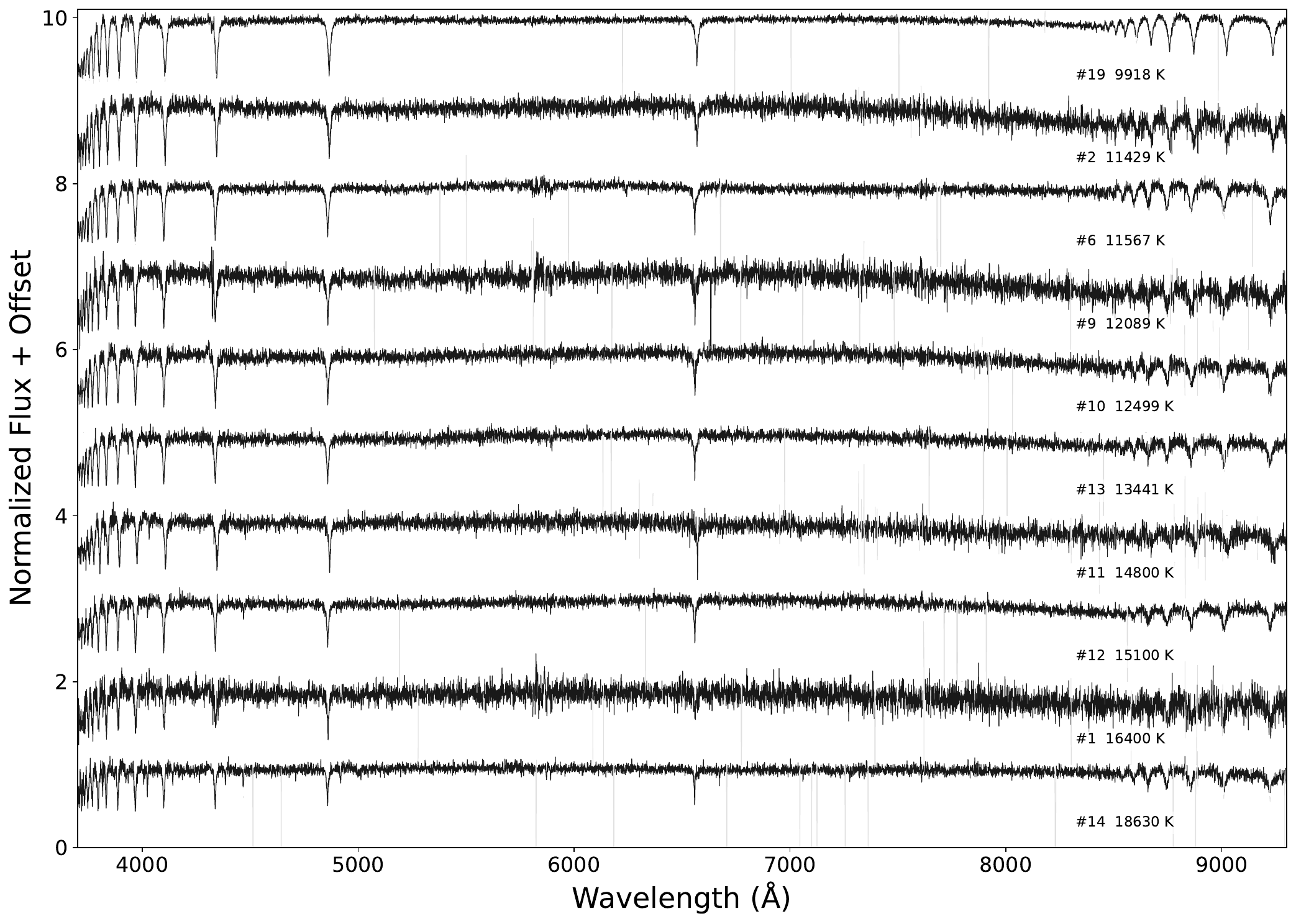}
     \caption{Normalized spectra of the 10 HVS candidates (includes BHB-HVS assuming MS nature) with galactocentric velocities higher than $600$ \kms\, from highest \teff\, (bottom) to lowest (top). The three arms of the DESI spectrograph are joined together to show the Balmer and Paschen series.}
     \label{fig:spectra}
 \end{figure*}

\section{Details of individual stars}
An interesting object in our initial sample is Gaia DR3 1328057527774422144. The star was a hot subdwarf candidate \citep{geiersdb} but was removed from the updated catalog of \citet{culpan}. It has previously been identified with photometric observations as a horizontal-branch star on the outskirts of the globular cluster M13 \citep{1955ApJ...122..171A}. It has a reliable \gaia\ parallax, corresponding to a distance of $2.5 \pm 0.6$ kpc, which, when combined with the SED, implies a stellar radius of $\sim0.2$ \rsun. Spectroscopic observations indicate a rapidly rotating star ($v\sin i > 80$ \kms), and comparison with MIST evolutionary tracks suggests a mass of $5.4 \pm 0.25$ \msun, implying a distance of $36.7 \pm 6$ kpc. The measured radial velocity of $-249 \pm 10$ \kms\, and proper motion components (-3.433, -2.274 mas/yr) are consistent with those of M13 \citep{2018MNRAS.478.1520B}. M13 is, however, known to lie at a distance of $\sim6.16$ kpc \citep{2022ApJ...934..150L}. This discrepancy indicates that both the \gaia\, parallax and the MIST-based inference are unreliable for this object, and that the star is most likely associated with M13.

Gaia DR3 4457139797898744832 shows evidence of being a metal-poor star (\feh$\sim-2$) with $v\sin i \sim 110$ \kms\,. The MIST tracks suggest a mass of $\sim1.36$ \msun\, at an age of $\sim 2.5$ Gyr, which leads us to believe that this star is a blue-straggler star in the halo.

\nolinenumbers
\FloatBarrier
\twocolumn
\onecolumn
\begin{landscape}
\newcounter{starindex}
\footnotesize
\setlength{\tabcolsep}{3pt}  

\tablefoot{Uncertainties on \logg\, and \teff\, include systematic and statistical uncertainties. Galactic Rest Frame velocity ($v_{\mathrm{GRF}}$) is for Model I of \citet{andreas}. Probability of being unbound ($P_{\mathrm{unbound}}$) is calculated from 1000 runs. Candidate column flags whether it HVS, HVS with low $v\sin i$, or MS runaway candidate. For BHB-HVS stars kinematic parameters are derived assuming MS nature, since they are not HVS candidates otherwise.}
\end{landscape}
\twocolumn
\FloatBarrier
\clearpage
\end{document}